\documentclass[journal]{IEEEtran}

\usepackage{amssymb,amsmath,amsfonts,amsthm,mathtools}
\usepackage{cite,graphicx}
\usepackage{float,subcaption}
\usepackage{booktabs}
\usepackage{tikz}
\usetikzlibrary{arrows.meta,positioning,calc,fit,backgrounds}
\usepackage{xcolor}
\usepackage{hyperref}
\hypersetup{colorlinks,breaklinks,citecolor=blue,linkcolor=blue,urlcolor=blue}
\usepackage[nameinlink,noabbrev,sort&compress]{cleveref}
\newtheorem{theorem}{Theorem}
\newtheorem{lemma}{Lemma}
\newtheorem{corollary}{Corollary}
\newtheorem{proposition}[theorem]{Proposition}
\newtheorem{problem}{Problem}
\theoremstyle{remark}
\newtheorem{remark}{Remark}
\newtheorem{assumption}{Assumption}
\theoremstyle{definition}

\Crefname{problem}{Problem}{Problems}
\Crefname{assumption}{Assumption}{Assumptions}
\newcommand{\abs}[1]{\lvert #1 \rvert}
\newcommand{\nrm}[1]{\lVert #1 \rVert}

\DeclareMathOperator{\diag}{diag}

\allowdisplaybreaks
\usepackage{newtxmath}
\usepackage{booktabs}
\usepackage{duckuments}

\begin{document}
\title{Admissibility-Preserving Control for Block Strict-Feedback Systems Under Time-Varying Asymmetric Actuator and State Constraints}
\author{Saurabh Kumar, \IEEEmembership{Member, IEEE}, Lohitvel Gopikannan, \IEEEmembership{Student Member, IEEE},\\ Shashi Ranjan Kumar, \IEEEmembership{Senior Member, IEEE}, and Abhinav Sinha, \IEEEmembership{Senior Member, IEEE}
\thanks{S. Kumar, L. Gopikannan and S.R. Kumar are with the Intelligent Systems and Control (ISaC) Lab, Department of Aerospace Engineering, Indian Institute of Technology Bombay, Maharashtra 400076 India (e-mails: \{saurabh.k,lohitvel,srk\}@aero.iitb.ac.in).}
\thanks{A. Sinha is with the Guidance, Autonomy, Learning, and Control for Intelligent Systems (GALACxIS) Lab, Department of Aerospace Engineering, University of Cincinnati, OH 45221, USA. (e-mail: abhinav.sinha@uc.edu).}}

\maketitle
\begin{abstract}
This paper addresses the tracking control of block strict-feedback nonlinear systems subject to time-varying asymmetric actuator and full state constraints. To that end, we develop a time-varying \emph{Admissibility-Preserving Control} (APC) framework, in which the \emph{Admissibility-Preserving Input Realization} (APIR) generates the plant input through a dynamic realization. The APIR guarantees input admissibility under the moving actuator set whenever each actuator bound contracts slower than the decay rate of the realization, and a closed-form lower bound on its gain is derived. The state constraints are handled by a logarithmic barrier coordinate at every backstepping step, with widths measured from the virtual control to the moving envelope of the corresponding state. We prove that every state lies inside its envelope, with each virtual control only required to remain inside the envelope of the next state, while the tracking error converges to zero exponentially. Simulations demonstrate the efficacy of the proposed safety-critical control scheme.
\end{abstract}


\section{Introduction}\label{sec:introduction}
\IEEEPARstart{I}{n} many practical control systems, actuator limits and state constraints vary over time owing to factors such as actuator derating, faults, and changing operating conditions. Although the design of feedback controllers for constrained nonlinear systems has attracted sustained research attention for over three decades, the systematic treatment of such time-varying constraints remains an open problem.  A foundational body of work addresses actuator limits through bounded-input stabilization, in which the feedback is constructed so that the control input remains within prescribed bounds. For linear plants, bounded globally stabilizing laws exist exactly when the plant is asymptotically null controllable \cite{doi:10.1109/9.362853}, while nested-saturation and low-gain designs achieve global stabilization of chains of integrators \cite{doi:10.1016/0167-6911(92)90001-9} and semiglobal stabilization of general null-controllable plants \cite{doi:10.1016/0167-6911(93)90033-3}. A comprehensive treatment of analysis and synthesis under actuator saturation is given in \cite{doi:10.1007/978-1-4612-0205-9}. Anti-windup designs follow a complementary approach, in which the nominal controller is retained, and an additional compensator modifies the closed-loop dynamics once the actuator saturates \cite{doi:10.1016/s0005-1098(02)00072-9,doi:10.1049/iet-cta:20070435}. For nonlinear plants, the saturation is usually absorbed by an auxiliary system or by an adaptive compensator placed around the nominal design \cite{doi:10.1109/TAC.2011.2122730,doi:10.1016/j.automatica.2011.01.025}. However, these designs assume time-invariant actuator bounds, which are mostly symmetric. Moreover, in the anti-windup and auxiliary-system designs, the correction is applied after the nominal design is complete. Consequently, the saturation limit is treated as a constant actuator parameter. In contrast, the fault-tolerant control literature studies authority that changes during operation, where the loss of effectiveness is usually modeled as a multiplicative gain to be estimated and compensated \cite{doi:10.1016/j.arcontrol.2008.03.008}.

Barrier Lyapunov functions (BLFs) provide an alternative mechanism for enforcing constraints in strict-feedback systems by adding a term to the Lyapunov function that diverges at the boundary. Consequently, the output/state remains inside its admissible set as long as the Lyapunov function remains bounded. In \cite{doi:10.1109/CDC.2005.1583507}, barrier functions were applied to systems with multiple state constraints, while the authors in \cite{doi:10.1016/j.automatica.2008.11.017} developed a BLF-based backstepping design for output-constrained strict-feedback systems, using both symmetric and asymmetric barriers. This framework was subsequently extended to partial state constraints \cite{doi:10.1080/00207179.2011.631192}, to full state constraints in pure-feedback systems \cite{doi:10.1016/j.automatica.2015.10.034} and under unknown control direction \cite{doi:10.1016/j.automatica.2016.10.011}, to output envelopes that vary in time \cite{doi:10.1016/j.automatica.2011.08.044}, to time-varying full state constraints through asymmetric barriers \cite{doi:10.1109/TSMC.2016.2633007}, and further to networked settings with moving state constraints \cite{doi:10.1109/TCYB.2016.2629268}.

However, in all of these designs, the barrier acts on the error between each state and its virtual control. Keeping this error bounded confines the state only if the virtual control itself remains within a reduced region of the state envelope. The resulting requirement is the feasibility condition, which is usually verified offline by searching for design parameters \cite{doi:10.1016/j.automatica.2015.10.034,doi:10.1109/TAC.2018.2845707}. Integral BLFs place the barrier on the state itself and reduce the conservatism of this condition while retaining it \cite{doi:10.1109/CDC.2012.6426196,doi:10.1109/TCYB.2016.2581173}. The authors in \cite{doi:10.1109/TAC.2018.2845707} removed the feasibility condition through a nonlinear state-dependent transformation combined with dynamic surface control. However, the tracking error in \cite{doi:10.1109/TAC.2018.2845707} is only shown to be uniformly ultimately bounded.

Moreover, prescribed-performance and funnel designs confine the tracking error to a prescribed envelope by transforming the constrained error into an unconstrained coordinate \cite{doi:10.1109/TAC.2008.929402,doi:10.1051/cocv:2002064}. Since the envelope is a design function of time, time-varying constraints on the tracking error are handled directly. However, these designs constrain only the tracking error, and the input constraints require an additional mechanism. In funnel control, the input constraints were treated as hard constraints in \cite{doi:10.1109/TAC.2024.3352362}, where the funnel boundary is widened by a dynamic law whenever the input saturates and may grow unbounded if the saturation persists. Within prescribed performance control, the authors in \cite{doi:10.1109/TAC.2023.3322619} handled the input constraints by modifying the reference trajectory, while \cite{doi:10.1109/LCSYS.2023.3281346,doi:10.1109/TSMC.2024.3462728} relaxed the performance specifications adaptively according to the actuator limits, with the boundedness of the closed-loop signals established only on compact sets unless the system is input-to-state stable. The authors in \cite{doi:10.1109/LCSYS.2022.3233723} retained both the performance and the input constraints, provided that a feasibility condition relating the prescribed performance to the input bounds holds. In these approaches, the input remains admissible either because the tracking objective is relaxed whenever the actuator authority is insufficient, or because the performance specification is restricted a priori.

Along similar lines, control barrier functions (CBFs) provide a mechanism for enforcing state constraints by rendering a safe set forward invariant. In \cite{doi:10.1109/TAC.2016.2638961}, CBFs were unified with control Lyapunov functions through a quadratic program (QP) solved online, which minimally modifies a nominal control law to retain safety. This framework was subsequently extended to constraints of high relative degree \cite{doi:10.1109/TAC.2021.3105491}. However, the input bounds enter the QP as additional constraints, and the QP may become infeasible when these bounds conflict with the CBF conditions \cite{doi:10.1109/TAC.2021.3105491}. The authors in \cite{doi:10.1109/TAC.2021.3074895} introduced adaptive CBFs that accommodate time-varying control bounds and guarantee the feasibility of the QP, provided that the original optimal control problem is feasible. Nevertheless, in these designs, the input bounds are satisfied only through the constraints of the online QP, so that input admissibility depends on the feasibility of the program at every time instant. Reference and command governors \cite{doi:10.1016/j.automatica.2016.08.013} also retain the nominal control law, and enforce the constraints by modifying the reference applied to the closed loop.

The Admissibility-Preserving Control (APC) framework has been recently introduced in our earlier work for a time-invariant asymmetric actuator set and a time-varying output constraint \cite{doi:10.48550/arXiv.2608.15375}. APC generates the physical input through a differential equation, termed the admissibility-preserving input realization (APIR), whose effective gain vanishes at the actuator bounds. Therein, the admissible input set is forward invariant by construction, and the realized input is continuously differentiable, which allows it to be treated as an additional state in a recursive Lyapunov design. This framework has since been extended to networked coordination \cite{doi:10.48550/arXiv.2609.09384}, and the constrained-motion applications that motivate it are described in \cite{doi:10.2514/1.G009445}.

However, three gaps remain when the admissible sets vary in time. First, the forward invariance of a time-invariant actuator set does not carry over to a moving one. Second, BLF backstepping \cite{doi:10.1080/00207179.2011.631192,doi:10.1016/j.automatica.2015.10.034,doi:10.1109/TSMC.2016.2633007} confines each state only if its virtual control keeps the barrier width away from the envelope, a feasibility condition that is verified offline \cite{doi:10.1016/j.automatica.2015.10.034,doi:10.1109/TAC.2018.2845707}. Third, since the command of the APIR is formed by dividing by the realization gain \cite{doi:10.48550/arXiv.2608.15375}, this gain cancels in the closed loop, and an explicit compatibility condition that bounds the command is required. To the best of our knowledge, tracking control of nonlinear systems under time-varying asymmetric actuator and full-state constraints, with both certified by a single explicit condition, remains unaddressed. We close this gap by extending APC to moving admissible sets for block strict-feedback systems. We summarize our contributions below.

\begin{itemize}
\item A time-varying asymmetric APIR is developed whose moving actuator set is forward invariant under bounded commands, provided that each actuator bound contracts slower than the decay rate of the realization. The analysis in a normalized coordinate also establishes a compact invariant interior set and a closed-form lower bound on the APIR gain.
\item A logarithmic barrier coordinate is applied at every block of the backstepping recursion, with its widths measured from the virtual control to the moving envelope of the corresponding state. Unlike BLF backstepping \cite{doi:10.1080/00207179.2011.631192,doi:10.1016/j.automatica.2015.10.034,doi:10.1109/TSMC.2016.2633007}, the full state constraints hold without barrier widths reserved for the errors, and the feasibility condition is replaced by a separation condition that only requires each virtual control to lie inside the envelope of the next state. This separation condition and the boundedness of the commanded input are combined into a single APC compatibility test on a Lyapunov sublevel set.
\item A closed-form ratio between each error and its barrier coordinate is derived, which carries the cross terms between consecutive coordinates. Consequently, the Lyapunov identity of the closed-loop system is exact and yields an explicit decay rate.
\item Under APC compatibility, we show that the tracking error converges exponentially to zero with an explicit rate and amplitude, while all state envelopes and the moving actuator set remain forward invariant and all closed-loop signals remain bounded. Moreover, provided that the APIR is designed on a correspondingly tightened actuator set that satisfies the boundary-rate condition, input admissibility transfers to the physical actuator when the applied input deviates from the realized input by a bounded error.
\end{itemize}
\section{Problem Formulation}\label{sec:problem}
Throughout this paper, $\mathbb{R}^{m}$ denotes the $m$-dimensional Euclidean space, $\mathbb{R}_{+}$ the set of positive reals, and $I$, $\mathbf{1}$, and $\mathbf{0}$ the identity matrix and vectors of ones and zeros, respectively, of appropriate dimensions. The transpose and absolute value are denoted by $(\cdot)^{\top}$ and $\abs{\cdot}$, respectively. Vector inequalities, products, and quotients are understood componentwise. For square matrices, $A\succeq B$ indicates that $A-B$ is positive semidefinite. The operators $\diag(\cdot)$ and $\sigma_{\min}(\cdot)$ denote a diagonal matrix and the smallest singular value, respectively. The Euclidean and infinity norms are denoted by $\nrm{\cdot}$ and $\nrm{\cdot}_{\infty}$, respectively, and $\operatorname{dist}_{\infty}(v,\mathbb{K})\coloneqq\inf_{w\in\mathbb{K}}\nrm{v-w}_{\infty}$ denotes the infinity-norm distance to $\mathbb{K}\subset\mathbb{R}^{m}$. A time-varying set $\mathbb{S}(t)$ is forward invariant if $x(0)\in\mathbb{S}(0)$ implies $x(t)\in\mathbb{S}(t)$ $\forall t\geq0$.

Consider the class of block strict-feedback nonlinear systems given by
\begin{subequations}\label{eq:plant}
\begin{align}
\dot{x}_{i} =&~f_{i}(\bar{x}_{i}) + G_{i}(\bar{x}_{i})x_{i+1}, \quad i=1,\ldots,n-1, \label{eq:plant_i}\\
\dot{x}_{n} =&~f_{n}(\bar{x}_{n}) + G_{n}(\bar{x}_{n})u, \label{eq:plant_n}\\
y =&~x_{1}, \label{eq:plant_y}
\end{align}
\end{subequations}
where $x_{i}\in\mathbb{R}^{m}$ is the $i^{\mathrm{th}}$ state block, $\bar{x}_{i}=[x_{1}^{\top},\ldots,x_{i}^{\top}]^{\top}$, $u\in\mathbb{R}^{m}$ is the input applied to the plant, $y\in\mathbb{R}^{m}$ is the measured output, and $f_{i}:\mathbb{R}^{im}\to\mathbb{R}^{m}$ and $G_{i}:\mathbb{R}^{im}\to\mathbb{R}^{m\times m}$ are smooth. Note that the dynamics of $x_{i}$ are affine in $x_{i+1}$, which acts as the input to the $i^{\mathrm{th}}$ subsystem. In other words, \eqref{eq:plant} is control affine at every step of the recursion.

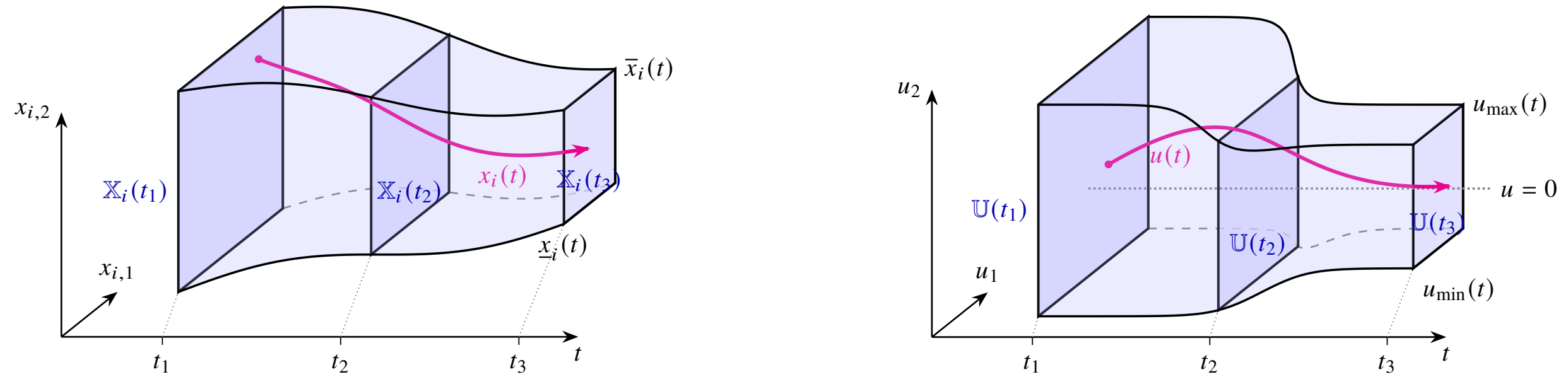
\begin{figure*}[!ht]
\centering
\tikzset{tube/.style={black,line width=0.8pt},
hid/.style={gray,dashed,line width=0.6pt},
face/.style={blue!30,fill opacity=0.25,draw=none},
slice/.style={black,line width=0.9pt,fill=blue!15,fill opacity=0.75},
tr/.style={magenta,line width=1.3pt},
ax/.style={-{Stealth[length=2mm]},line width=0.6pt},
lab/.style={blue!70!black,fill=none,fill opacity=0.8,text opacity=1,inner sep=1pt}}

\begin{subfigure}{0.49\linewidth}
\centering
\resizebox{0.8\linewidth}{!}{%
\begin{tikzpicture}[x={(0.74cm,0cm)},y={(0.5cm,0.4cm)},z={(0cm,1cm)},font=\small,
declare function={alo(\t)=-1.2+0.12*\t; ahi(\t)=1.4-0.1*\t;
blo(\t)=-1.0+0.1*\t+0.15*sin(\t r); bhi(\t)=1.5-0.08*\t+0.15*sin(\t r);
xa(\t)=0.8*exp(-0.35*\t)+0.1*sin(1.2*\t r); xb(\t)=0.9*exp(-0.4*\t)+0.2+0.15*sin(\t r);}]
\draw[ax] (-1.7,-1.6,-1.4) -- (7.0,-1.6,-1.4) node[below] {$t$};
\draw[ax] (-1.7,-1.6,-1.4) -- (-1.7,-0.2,-1.4) node[above] {$x_{i,1}$};
\draw[ax] (-1.7,-1.6,-1.4) -- (-1.7,-1.6,1.4) node[left] {$x_{i,2}$};
\foreach \T/\L in {0/1,3/2,6/3}{
\draw[gray,densely dotted] (\T,-1.6,-1.4) -- (\T,{alo(\T)},{blo(\T)});
\draw (\T,-1.6,-1.4) -- ++(0,0,-0.1) node[below] {$t_{\L}$};}
\draw[hid] plot[domain=0:6,samples=40,smooth] (\x,{ahi(\x)},{blo(\x)});
\foreach \T in {0,3,6}{
\draw[slice] (\T,{alo(\T)},{blo(\T)}) -- (\T,{ahi(\T)},{blo(\T)}) -- (\T,{ahi(\T)},{bhi(\T)}) -- (\T,{alo(\T)},{bhi(\T)}) -- cycle;
}
\draw[tr,-{Stealth[length=2.2mm]}] plot[domain=0:6,samples=60,smooth] (\x,{xa(\x)},{xb(\x)});
\fill[magenta] (0,{xa(0)},{xb(0)}) circle (1.4pt);
\node[magenta,below] at (4.6,{xa(4.6)},{xb(4.6)}) {$x_{i}(t)$};
\fill[face] plot[domain=0:6,samples=40,smooth] (\x,{alo(\x)},{blo(\x)}) -- plot[domain=6:0,samples=40,smooth] (\x,{alo(\x)},{bhi(\x)}) -- cycle;
\fill[face] plot[domain=0:6,samples=40,smooth] (\x,{alo(\x)},{bhi(\x)}) -- plot[domain=6:0,samples=40,smooth] (\x,{ahi(\x)},{bhi(\x)}) -- cycle;
\foreach \A/\B in {alo/blo,alo/bhi,ahi/bhi}
\draw[tube] plot[domain=0:6,samples=40,smooth] (\x,{\A(\x)},{\B(\x)});
\node[right] at (6,{ahi(6)},{bhi(6)}) {$\overline{x}_{i}(t)$};
\node[below] at (6,{alo(6)},{blo(6)}) {$\underline{x}_{i}(t)$};
\node[blue!70!black,left] at (0,{alo(0)},{0.5*blo(0)+0.5*bhi(0)}) {$\mathbb{X}_{i}(t_{1})$};
\foreach \T/\L in {3/2,6/3}
\node[lab] at (\T,{0.5*alo(\T)+0.5*ahi(\T)},{0.8*blo(\T)+0.2*bhi(\T)}) {$\mathbb{X}_{i}(t_{\L})$};
\end{tikzpicture}}
\caption{State set.}
\label{fig:boxes_x}
\end{subfigure}%
\hfill
\begin{subfigure}{0.49\linewidth}
\centering
\resizebox{0.8\linewidth}{!}{%
\begin{tikzpicture}[x={(0.74cm,0cm)},y={(0.5cm,0.4cm)},z={(0cm,1cm)},font=\small,
declare function={ulo(\t)=-1.0+0.25*tanh(1.6*(\t-3.6)); uhi(\t)=1.0+0.5*tanh(2*(2.8-\t));
vlo(\t)=-0.9+0.2*tanh(1.6*(\t-3.6)); vhi(\t)=1.2+0.35*tanh(2*(2.8-\t));
ua(\t)=0.2+0.6*exp(-0.3*(\t-1.5)^2); ub(\t)=0.1+0.5*sin(0.8*\t r)*exp(-0.2*\t);}]
\draw[ax] (-1.7,-1.4,-1.3) -- (7.0,-1.4,-1.3) node[below] {$t$};
\draw[ax] (-1.7,-1.4,-1.3) -- (-1.7,0.0,-1.3) node[above] {$u_{1}$};
\draw[ax] (-1.7,-1.4,-1.3) -- (-1.7,-1.4,1.8) node[left] {$u_{2}$};
\foreach \T/\L in {0/1,3/2,6/3}{
\draw[gray,densely dotted] (\T,-1.4,-1.3) -- (\T,{ulo(\T)},{vlo(\T)});
\draw (\T,-1.4,-1.3) -- ++(0,0,-0.1) node[below] {$t_{\L}$};}
\draw[hid] plot[domain=0:6,samples=40,smooth] (\x,{uhi(\x)},{vlo(\x)});
\foreach \T in {0,3,6}{
\draw[slice] (\T,{ulo(\T)},{vlo(\T)}) -- (\T,{uhi(\T)},{vlo(\T)}) -- (\T,{uhi(\T)},{vhi(\T)}) -- (\T,{ulo(\T)},{vhi(\T)}) -- cycle;
}
\draw[gray,densely dotted,line width=0.8pt] (0,0,0) -- (6.8,0,0) node[right,black] {$u=0$};
\draw[tr,-{Stealth[length=2.2mm]}] plot[domain=0:6,samples=60,smooth] (\x,{ua(\x)},{ub(\x)});
\fill[magenta] (0,{ua(0)},{ub(0)}) circle (1.4pt);
\node[magenta,below] at (0.9,{ua(0.9)},{ub(0.9)}) {$u(t)$};
\fill[face] plot[domain=0:6,samples=40,smooth] (\x,{ulo(\x)},{vlo(\x)}) -- plot[domain=6:0,samples=40,smooth] (\x,{ulo(\x)},{vhi(\x)}) -- cycle;
\fill[face] plot[domain=0:6,samples=40,smooth] (\x,{ulo(\x)},{vhi(\x)}) -- plot[domain=6:0,samples=40,smooth] (\x,{uhi(\x)},{vhi(\x)}) -- cycle;
\foreach \A/\B in {ulo/vlo,ulo/vhi,uhi/vhi}
\draw[tube] plot[domain=0:6,samples=40,smooth] (\x,{\A(\x)},{\B(\x)});
\node[right] at (6,{uhi(6)},{vhi(6)}) {$u_{\max}(t)$};
\node[below right] at (6,{ulo(6)},{vlo(6)}) {$u_{\min}(t)$};
\node[blue!70!black,left] at (0,{ulo(0)},{0.5*vlo(0)+0.5*vhi(0)}) {$\mathbb{U}(t_{1})$};
\foreach \T/\L in {3/2,6/3}
\node[lab] at (\T,{0.5*ulo(\T)+0.5*uhi(\T)},{0.8*vlo(\T)+0.2*vhi(\T)}) {$\mathbb{U}(t_{\L})$};
\end{tikzpicture}}
\caption{Input set.}
\label{fig:boxes_u}
\end{subfigure}
\caption{Time-varying state and input sets.}
\label{fig:boxes}
\end{figure*}

For a safe operation, each state block $x_{i}$ and the plant input $u$ are required to remain inside the time-varying sets
\begin{align}
\mathbb{X}_{i}(t) &\coloneqq \left\{x\in\mathbb{R}^{m}\mid \underline{x}_{i}(t)<x<\overline{x}_{i}(t)\right\}, i=1,\ldots,n, \label{eq:state_set}\\
\mathbb{U}(t) &\coloneqq \left\{u\in\mathbb{R}^{m}\mid u_{\min}(t)<u<u_{\max}(t)\right\}, \label{eq:input_set}
\end{align}
with $u_{\min}(t)<0<u_{\max}(t)$ for all $t\geq0$. Since $y=x_{1}$, $\mathbb{X}_{1}(t)$ also serves as the output constraint. The sets are illustrated for $m=2$ in \Cref{fig:boxes}.

\begin{assumption}[Model regularity and control effectiveness]\label{ass:model}
For each $i$, the maps $f_{i}$ and $G_{i}$ are known, smooth, and available for synthesis, and $G_{i}(\bar{x}_{i})$ is invertible with $\sigma_{\min}(G_{i}(\bar{x}_{i}))\geq\underline{g}_{i}>0$ for all $\bar{x}_{i}$, where $\underline{g}_{i}$ is a known constant.
\end{assumption}
The bound $\underline{g}_{i}$ keeps every block uniformly invertible, so no direction of $x_{i}$ loses actuation through $x_{i+1}$. No upper bound is imposed, admitting input gains that grow with the state.

\begin{assumption}\label{ass:ref}
The reference $y_{d}$ is $n+1$ times continuously differentiable with bounded derivatives, and there exists $\varepsilon_{y}>0$ such that $\underline{x}_{1}(t)+\varepsilon_{y}\mathbf{1}\leq y_{d}(t)\leq\overline{x}_{1}(t)-\varepsilon_{y}\mathbf{1}$ for all $t\geq0$.
\end{assumption}
Owing to $\varepsilon_{y}$, the reference lies strictly inside the envelope, and hence exact tracking itself is an admissible motion.

\begin{assumption}[State envelopes]\label{ass:env}
The bounds $\underline{x}_{1}$ and $\overline{x}_{1}$ are $n+1$ times continuously differentiable, and for $i=2,\ldots,n$ the bounds $\underline{x}_{i}$ and $\overline{x}_{i}$ are $n+2-i$ times continuously differentiable. All of these derivatives are bounded, the bounds themselves are bounded, and $\underline{x}_{i}(t)<\overline{x}_{i}(t)$ for all $t\geq0$ and every $i$.
\end{assumption}
Envelopes generated by a scheduling law or by a reference model are smooth functions of time with bounded derivatives and satisfy \Cref{ass:env}. The strict inequality $\underline{x}_{i}(t)<\overline{x}_{i}(t)$ keeps every set $\mathbb{X}_{i}(t)$ nonempty for all $t\geq0$. We are now in a position to formally define our problem statement.
\begin{problem}[Admissibility-preserving constrained tracking]\label{prob:1}
Consider the system \eqref{eq:plant} with the time-varying sets \eqref{eq:state_set} and \eqref{eq:input_set}. Develop a control law such that, for every initial condition with $x_{i}(0)\in\mathbb{X}_{i}(0)$, $i=1,\ldots,n$, and $u(0)\in\mathbb{U}(0)$ that is compatible with the available actuator authority, the following hold for all $t\geq0$:
\begin{enumerate}
\item every state block satisfies $x_{i}(t)\in\mathbb{X}_{i}(t)$, $i=1,\ldots,n$;
\item the realized plant input satisfies $u(t)\in\mathbb{U}(t)$;
\item the tracking error $y(t)-y_{d}(t)$ converges to zero exponentially, and all closed-loop signals remain bounded.
\end{enumerate}
\end{problem}
Note that \Cref{prob:1} is posed on compatible initial conditions and not globally, since, with finite authority, an initial state far from the reference, or a demanding reference, cannot be steered by any controller without violating \eqref{eq:state_set} or \eqref{eq:input_set}. Moreover, the first requirement is stronger than an output constraint, since every intermediate state block must remain inside its moving envelope. Such a requirement arises, for instance, in flight vehicles, where the angle of attack and body rates must remain within envelopes that vary with flight conditions \cite{altunkaya2026aircraft,doi:10.2514/1.G008792}, and in robotic manipulators \cite{9838184}, where joint velocities are limited in addition to joint positions.

\section{The Time-Varying APC Framework}\label{sec:main}
This section develops time-varying APC through three elements: \emph{APIR realizes admissibility under a moving actuator set}, \emph{APC synthesis achieves the control objective around the admissible realization}, and \emph{APC compatibility certifies feasible use of finite authority}. We first establish the admissibility and gain properties of the APIR under moving bounds. The realization is then incorporated into backstepping, with a barrier coordinate transformation at every block so that each state remains inside its moving envelope. Finally, the APC compatibility condition is stated, under which every constraint is met with the available actuator authority.

\subsection{Admissibility-Preserving Input Realization under Moving Bounds}\label{subsec:apir}
The APIR is first analyzed as an independent dynamic module, whose admissibility depends only on the boundedness of the commanded input and on the rate at which the actuator bounds move, and not on the plant. In this paper, the APIR of \cite{doi:10.48550/arXiv.2608.15375} is extended to the moving actuator set \eqref{eq:input_set} through the componentwise realization
\begin{equation}\label{eq:apir}
\dot{u} = \mathcal{G}(u,t)u_{c}-\mathcal{F}(u),
\end{equation}
where $u_{c}$ is the commanded input generated by the synthesis layer, $u$ is the realized actuator output supplied to the plant, and $\mathcal{G}(u,t)$ and $\mathcal{F}(u)$ are given by
\begin{equation}\label{eq:GF}
\mathcal{G}(u,t)\coloneqq p_{1}\diag\left(\mathcal{S}_{l}(u_{l},t)\right),
\mathcal{F}(u)\coloneqq p_{1}p_{2}u,
\end{equation}
\begin{equation}\label{eq:S}
\mathcal{S}_{l}(u_{l},t)\coloneqq
\begin{cases}
1-\left(\dfrac{u_{l}}{u_{\max,l}(t)}\right)^{\gamma}, & u_{l}>0,\\[8pt]
1-\left(\dfrac{u_{l}}{u_{\min,l}(t)}\right)^{\gamma}, & u_{l}\leq0,
\end{cases}
\end{equation}
where $p_{1},p_{2}\in\mathbb{R}_{+}$ are design parameters, $\gamma$ is an even positive integer that determines the smoothness order and transition profile of the realization, and $u_{\min,l}(t)<0<u_{\max,l}(t)$ are the lower and upper bounds of the $l^{\mathrm{th}}$ actuator channel. Since both branches of \eqref{eq:S} yield $\mathcal{S}_{l}=1$ at $u_{l}=0$, the APIR in \eqref{eq:apir} is locally Lipschitz in $u$. In contrast to \cite{doi:10.48550/arXiv.2608.15375}, the bounds are time-varying, and hence $\mathcal{G}$ depends explicitly on time. A parametric sensitivity study of the APIR is reported in \cite{doi:10.48550/arXiv.2608.15375}.

\begin{assumption}[Admissible actuator bounds and boundary rates]\label{ass:actuator}
The bounds $u_{\max}(t)$ and $u_{\min}(t)$ are continuously differentiable, and there exist constants $\underline{u}_{\max}>0$, $\underline{u}_{\min}>0$, and $\varepsilon_{0}\in(0,p_{1}p_{2})$ such that, for every component $l$ and all $t\geq0$,
\begin{align}\label{eq:rate}
u_{\max,l}(t)&\geq\underline{u}_{\max}, u_{\min,l}(t)\leq-\underline{u}_{\min},\\
\frac{\dot{u}_{\max,l}(t)}{u_{\max,l}(t)}&\geq-p_{1}p_{2}+\varepsilon_{0},
\frac{\dot{u}_{\min,l}(t)}{u_{\min,l}(t)}\geq-p_{1}p_{2}+\varepsilon_{0}. \label{eq:rate2}
\end{align}
\end{assumption}
Note that \eqref{eq:rate2} constrains the logarithmic derivatives of $u_{\max,l}$ and $\abs{u_{\min,l}}$, that is, their instantaneous exponential rates of change. Therefore, \eqref{eq:rate2} allows each actuator boundary to contract only at an exponential rate below $p_{1}p_{2}$, which is the rate at which the term $\mathcal{F}(u)=p_{1}p_{2}u$ in \eqref{eq:GF} pulls the realized input toward the origin, whereas expansion is in general unrestricted. The margin $\varepsilon_{0}$ must be uniform in time, since a pointwise strict inequality admits ratios that approach $-p_{1}p_{2}$ asymptotically, under which the confinement margin degenerates. The condition in \eqref{eq:rate} ensures that the available authority does not vanish asymptotically, which keeps the ratio between the command bound and the available authority uniformly bounded.
\begin{theorem}[Intrinsic admissibility and compact interiority under moving bounds]\label{thm:apir}
Consider \eqref{eq:apir} under \Cref{ass:actuator}, with a piecewise-continuous command satisfying $\nrm{u_{c}(t)}_{\infty}\leq\xi$ for all $t\geq0$. Define
\begin{equation}\label{eq:K}
K_{\max}\coloneqq\frac{p_{1}\xi}{\underline{u}_{\max}},\,
K_{\min}\coloneqq\frac{p_{1}\xi}{\underline{u}_{\min}},
\end{equation}
and let $s^{*},\zeta^{*}\in(0,1)$ satisfy
\begin{equation}\label{eq:zstar}
K_{\max}\left[1-(s^{*})^{\gamma}\right]<s^{*}\varepsilon_{0},\,
K_{\min}\left[1-(\zeta^{*})^{\gamma}\right]<\zeta^{*}\varepsilon_{0}.
\end{equation}
For every $u(0)\in\mathbb{U}(0)$, define
\begin{align}\label{eq:zbar}
\bar{s}&\coloneqq\max\left\{s^{*},\ \max_{l}\frac{u_{l}(0)}{u_{\max,l}(0)}\right\},\nonumber\\
\bar{\zeta}&\coloneqq\max\left\{\zeta^{*},\ \max_{l}\frac{u_{l}(0)}{u_{\min,l}(0)}\right\}.
\end{align}
Then $\bar{s},\bar{\zeta}\in(0,1)$, and the solution of \eqref{eq:apir} exists for all $t\geq0$ and satisfies
\begin{equation}\label{eq:confine}
\bar{\zeta}u_{\min}(t)\leq u(t)\leq\bar{s}u_{\max}(t),\qquad t\geq0,
\end{equation}
Consequently, $\mathbb{U}(t)$ is forward invariant, and the realization gain satisfies
\begin{equation}\label{eq:gainbound}
\mathcal{G}(u(t),t)\succeq\underline{\mathcal{G}}I,\,
\underline{\mathcal{G}}\coloneqq p_{1}\left[1-\max\{\bar{s},\bar{\zeta}\}^{\gamma}\right]>0 .
\end{equation}
\end{theorem}

\begin{proof}
Note that $s^{*}$ and $\zeta^{*}$ exist, since, as $s^{*}\to1$ and $\zeta^{*}\to1$, the left sides of \eqref{eq:zstar} tend to zero, whereas the right sides tend to $\varepsilon_{0}>0$. As \eqref{eq:apir} is diagonal, it suffices to consider a single component, whose index is suppressed. Local existence and uniqueness follow from the local Lipschitz continuity of \eqref{eq:apir} in $u$ and its piecewise continuity in $t$. Define the normalized coordinate $s\coloneqq u/u_{\max}$, which is well defined by \eqref{eq:rate}. For $u>0$, one has $\mathcal{S}=1-s^{\gamma}$ and $u=su_{\max}$, so differentiating $s$ along \eqref{eq:apir} yields
\begin{align}
\dot{s}&=\frac{\dot{u}}{u_{\max}}-s\frac{\dot{u}_{\max}}{u_{\max}}=\frac{p_{1}\left[1-s^{\gamma}\right]u_{c}-p_{1}p_{2}su_{\max}}{u_{\max}}-s\frac{\dot{u}_{\max}}{u_{\max}}\nonumber\\
&=\frac{p_{1}u_{c}}{u_{\max}}\left[1-s^{\gamma}\right]-s\lambda,\label{eq:zdyn}
\end{align}
where $\lambda\coloneqq p_{1}p_{2}+\dot{u}_{\max}/u_{\max}$ satisfies $\lambda(t)\geq\varepsilon_{0}$ for all $t\geq0$ by \eqref{eq:rate2}. Since $1-s^{\gamma}>0$ for $s\in(0,1)$, $u_{c}\leq\xi$, and $u_{\max}\geq\underline{u}_{\max}$ by \eqref{eq:rate}, the first term in \eqref{eq:zdyn} is bounded above by $K_{\max}[1-s^{\gamma}]$. Together with $\lambda\geq\varepsilon_{0}$, this yields
\begin{equation}\label{eq:zbound}
\dot{s}\leq K_{\max}\left[1-s^{\gamma}\right]-s\varepsilon_{0},\quad s\in(0,1).
\end{equation}
Now, let $s\in[s^{*},1)$. As $s\mapsto1-s^{\gamma}$ is decreasing on $[0,1)$, one has $K_{\max}[1-s^{\gamma}]\leq K_{\max}[1-(s^{*})^{\gamma}]$. Moreover, \eqref{eq:zstar} gives $K_{\max}[1-(s^{*})^{\gamma}]<s^{*}\varepsilon_{0}$, and $s^{*}\varepsilon_{0}\leq s\varepsilon_{0}$ since $s\geq s^{*}$. Combining these inequalities yields $K_{\max}[1-s^{\gamma}]<s\varepsilon_{0}$, so the right side of \eqref{eq:zbound} is negative. Therefore, $\dot{s}<0$ for all $s\in[s^{*},1)$.

Since $\bar{s}\in[s^{*},1)$, one has $\dot{s}<0$ whenever $s=\bar{s}$, so $s$ cannot cross $\bar{s}$ from below. As $s(0)\leq\bar{s}$ by \eqref{eq:zbar}, it follows that $s(t)\leq\bar{s}$ for all $t\geq0$. Multiplying by $u_{\max}(t)>0$ yields the upper inequality in \eqref{eq:confine}.

The lower inequality follows similarly. Define $\zeta\coloneqq u/u_{\min}$, which is well defined by \eqref{eq:rate} and positive exactly when $u<0$. For $u<0$, one has $\mathcal{S}=1-\zeta^{\gamma}$ and $u=\zeta u_{\min}$, so differentiating $\zeta$ along \eqref{eq:apir} yields $\dot{\zeta}=(p_{1}u_{c}/u_{\min})[1-\zeta^{\gamma}]-\zeta(p_{1}p_{2}+\dot{u}_{\min}/u_{\min})$,
where $p_{1}p_{2}+\dot{u}_{\min}/u_{\min}\geq\varepsilon_{0}$ for all $t\geq0$ by \eqref{eq:rate2}. Since $1-\zeta^{\gamma}>0$ for $\zeta\in(0,1)$, $u_{c}/u_{\min}\leq\xi/\abs{u_{\min}}$ because $u_{c}\geq-\xi$ and $u_{\min}<0$, and $\abs{u_{\min}}\geq\underline{u}_{\min}$ by \eqref{eq:rate}, the first term is bounded above by $K_{\min}[1-\zeta^{\gamma}]$. Together with \eqref{eq:rate2}, this yields $\dot{\zeta}\leq K_{\min}[1-\zeta^{\gamma}]-\zeta\varepsilon_{0}$ for $\zeta\in(0,1)$. For $\zeta\in[\zeta^{*},1)$, since $\zeta\mapsto1-\zeta^{\gamma}$ is decreasing and $\zeta\geq\zeta^{*}$, \eqref{eq:zstar} gives $K_{\min}[1-\zeta^{\gamma}]\leq K_{\min}[1-(\zeta^{*})^{\gamma}]<\zeta^{*}\varepsilon_{0}\leq\zeta\varepsilon_{0}$, so $\dot{\zeta}<0$. Since $\bar{\zeta}\in[\zeta^{*},1)$, $\zeta$ cannot cross $\bar{\zeta}$ from below. As $\zeta(0)\leq\bar{\zeta}$ by \eqref{eq:zbar}, it follows that $\zeta(t)\leq\bar{\zeta}$ for all $t\geq0$. Multiplying by $u_{\min}(t)<0$ yields lower inequality in \eqref{eq:confine}.

Finally, by \eqref{eq:GF} and \eqref{eq:S}, each diagonal entry of $\mathcal{G}$ equals $p_{1}[1-s^{\gamma}]$ when $u>0$ and $p_{1}[1-\zeta^{\gamma}]$ when $u\leq0$. Since $0<s\leq\bar{s}$ in the first case and $0\leq\zeta\leq\bar{\zeta}$ in the second, the entry is at least $p_{1}[1-\max\{\bar{s},\bar{\zeta}\}^{\gamma}]$, which gives \eqref{eq:gainbound}. Since $u$ remains bounded, local Lipschitz continuity yields forward completeness.
\end{proof}

\begin{remark}\label{rem:gain}
Note that the normalized coordinates $s$ and $\zeta$ also yield the gain bound \eqref{eq:gainbound} in closed form. In \cite{doi:10.48550/arXiv.2608.15375}, the corresponding bound for fixed actuator limits was obtained by minimizing a continuous function over a compact invariant interval, so the resulting constant was not available explicitly. In contrast, $\underline{\mathcal{G}}$ in \eqref{eq:gainbound} follows directly from $\bar{s}$ and $\bar{\zeta}$, which are computed from \eqref{eq:zstar} and \eqref{eq:zbar}. In addition, \eqref{eq:rate2} can be understood as a rate budget for authority loss, so the product $p_{1}p_{2}$ sets both the equilibrium margin of the realization and the fastest admissible contraction of the actuator bounds.
\end{remark}

\begin{remark}\label{rem:gamma2}
Let $\gamma=2$, and recall $K_{\max}$ and $K_{\min}$ from \eqref{eq:K} and the rate margin $\varepsilon_{0}$ from \eqref{eq:rate2}. Then the first condition in \eqref{eq:zstar} is equivalent to $K_{\max}(s^{*})^{2}+\varepsilon_{0}s^{*}-K_{\max}>0$. The left side equals $-K_{\max}<0$ at $s^{*}=0$ and $\varepsilon_{0}>0$ at $s^{*}=1$, so this quadratic has exactly one root in $(0,1)$, given by
\begin{equation}\label{eq:splus}
s_{+}\coloneqq\frac{\sqrt{\varepsilon_{0}^{2}+4K_{\max}^{2}}-\varepsilon_{0}}{2K_{\max}},
\end{equation}
and the condition holds if and only if $s^{*}\in(s_{+},1)$. Similarly, the second condition in \eqref{eq:zstar} holds if and only if $\zeta^{*}\in(\zeta_{+},1)$, where $\zeta_{+}$ is given by \eqref{eq:splus} with $K_{\min}$ in place of $K_{\max}$. Hence, for $\gamma=2$, the constants $\bar{s}$, $\bar{\zeta}$, and $\underline{\mathcal{G}}$ in \eqref{eq:zbar} and \eqref{eq:gainbound} are available in closed form.
\end{remark}

\subsection{APIR-Augmented APC for State-Constrained Tracking}\label{subsec:coords}
We now interconnect the APIR \eqref{eq:apir} with the plant \eqref{eq:plant} to solve \Cref{prob:1}. To enforce the state constraints, we apply a smooth asymmetric logarithmic barrier coordinate at every block, measured from a reference signal to the lower and upper boundaries of the corresponding envelope. At the first block, the reference is $y_{d}$, whereas at block $i\geq2$, it is the virtual control that the preceding step assigns to $x_{i}$. Now, the barrier widths become the distances from that virtual control to the envelope of $x_{i}$, and a finite coordinate is equivalent to admissibility of the state itself.

To that end, we define the errors
\begin{equation}\label{eq:errors}
\varphi_{1}\coloneqq x_{1}-y_{d},\,
\varphi_{i}\coloneqq x_{i}-\eta_{i-1},\,
\varrho\coloneqq u-\eta_{n},
\end{equation}
for $i=2,\ldots,n$, where $\eta_{i}$, $i=1,\ldots,n-1$, is the virtual control input for $x_{i+1}$, and $\eta_{n}$ is the virtual input demanded of the APIR. The corresponding barrier widths are defined as
\begin{align}
\alpha_{1}&\coloneqq y_{d}-\underline{x}_{1},\, \beta_{1}\coloneqq\overline{x}_{1}-y_{d},\label{eq:ab1}\\
\alpha_{i}&\coloneqq\eta_{i-1}-\underline{x}_{i},\, \beta_{i}\coloneqq\overline{x}_{i}-\eta_{i-1},\, i=2,\ldots,n.\label{eq:abi}
\end{align}
Note that $\alpha_{1}\succ0$ and $\beta_{1}\succ0$ by \Cref{ass:ref}. For $i=1,\ldots,n$, a componentwise barrier coordinate transformation is defined as
\begin{equation}\label{eq:zi}
z_{i,l}\coloneqq\ln\!\left(\frac{\beta_{i,l}[\alpha_{i,l}+\varphi_{i,l}]}{\alpha_{i,l}[\beta_{i,l}-\varphi_{i,l}]}\right).
\end{equation}
Since the widths $\alpha_{i,l}$ and $\beta_{i,l}$ vary with time, $z_{i,l}$ changes both through the error $\varphi_{i,l}$ and through the widths. Writing \eqref{eq:zi} as $z_{i,l}=\ln\beta_{i,l}-\ln\alpha_{i,l}+\ln(\alpha_{i,l}+\varphi_{i,l})-\ln(\beta_{i,l}-\varphi_{i,l})$ and differentiating it, with the indices suppressed, gives $\dot{z}=\dot{\beta}/\beta-\dot{\alpha}/\alpha+(\dot{\alpha}+\dot{\varphi})/(\alpha+\varphi)-(\dot{\beta}-\dot{\varphi})/(\beta-\varphi)$, which, on collecting terms, becomes $\dot{z}_{i,l}=q_{i,l}\dot{\varphi}_{i,l}+\psi_{i,l}$, where
where
\begin{align}
q_{i,l}&\coloneqq\frac{\partial z_{i,l}}{\partial\varphi_{i,l}}=\frac{1}{\alpha_{i,l}+\varphi_{i,l}}+\frac{1}{\beta_{i,l}-\varphi_{i,l}}>0,\label{eq:q}\\
\psi_{i,l}&\coloneqq\dot{\alpha}_{i,l}\left(\frac{1}{\alpha_{i,l}+\varphi_{i,l}}-\frac{1}{\alpha_{i,l}}\right)+\dot{\beta}_{i,l}\left(\frac{1}{\beta_{i,l}}-\frac{1}{\beta_{i,l}-\varphi_{i,l}}\right).\label{eq:psi}
\end{align}
Here, $q_{i,l}$ is the rate at which $z_{i,l}$ changes with the error $\varphi_{i,l}$ when the widths are held fixed, whereas $\psi_{i,l}$ accounts for the effect of the moving widths and vanishes when $\dot{\alpha}_{i,l}=\dot{\beta}_{i,l}=0$. Stacking the components with $Q_{i}\coloneqq\diag(q_{i,l})$ and $\psi_{i}\coloneqq[\psi_{i,1},\ldots,\psi_{i,m}]^{\top}$ gives
\begin{equation}\label{eq:zdot}
\dot{z}_{i}=Q_{i}\dot{\varphi}_{i}+\psi_{i}.
\end{equation}
For $i\geq2$, the widths move with the virtual control, since $\dot{\alpha}_{i}=\dot{\eta}_{i-1}-\dot{\underline{x}}_{i}$ and $\dot{\beta}_{i}=\dot{\overline{x}}_{i}-\dot{\eta}_{i-1}$. Moreover, $\alpha_{i}+\beta_{i}=2r_{i}$, where $\check{x}_{i}\coloneqq\tfrac12[\overline{x}_{i}+\underline{x}_{i}]$ and $r_{i}\coloneqq\tfrac12[\overline{x}_{i}-\underline{x}_{i}]$ denote the centre and the half-width of $\mathbb{X}_{i}(t)$.

\begin{lemma}\label{lem:coord}
Consider the barrier coordinate transformation \eqref{eq:zi} with $\alpha_{i}(t)\succ0$ and $\beta_{i}(t)\succ0$. Then the map $\varphi_{i}\mapsto z_{i}$ is a smooth diffeomorphism from $(-\alpha_{i},\beta_{i})$ onto $\mathbb{R}^{m}$, and $z_{i}=0$ if and only if $\varphi_{i}=0$. Moreover, $z_{i}$ is finite if and only if $\underline{x}_{i}(t)<x_{i}(t)<\overline{x}_{i}(t)$. Furthermore, the ratio $\kappa_{i,l}\coloneqq\varphi_{i,l}/z_{i,l}$ is given by
\begin{equation}\label{eq:kappa}
\kappa_{i,l}=\frac{\alpha_{i,l}\beta_{i,l}(e^{z_{i,l}}-1)}{z_{i,l}(\beta_{i,l}+\alpha_{i,l}e^{z_{i,l}})},
\end{equation}
which extends smoothly to $z_{i,l}=0$ with the value $\alpha_{i,l}\beta_{i,l}/(\alpha_{i,l}+\beta_{i,l})$, and $\kappa_{i,l}>0$ for all $z_{i,l}\in\mathbb{R}$.
\end{lemma}

\begin{proof}
Since \eqref{eq:zi} acts componentwise, it suffices to fix $t$ and $l$ and to write $\alpha$, $\beta$, $\varphi$ and $z$ for $\alpha_{i,l}(t)$, $\beta_{i,l}(t)$, $\varphi_{i,l}$ and $z_{i,l}$. On the interval $(-\alpha,\beta)$ both $\alpha+\varphi$ and $\beta-\varphi$ are positive, so \eqref{eq:zi} can be written as $z=\ln\beta-\ln\alpha+\ln(\alpha+\varphi)-\ln(\beta-\varphi),$
which is smooth in $\varphi$, and its derivative is $q_{i,l}>0$ from \eqref{eq:q}. The map $\varphi\mapsto z$ is strictly increasing and hence injective. As $\varphi\to-\alpha^{+}$, the term $\ln(\alpha+\varphi)$ tends to $-\infty$ while the remaining terms stay bounded, so $z\to-\infty$. Similarly, as $\varphi\to\beta^{-}$, the term $-\ln(\beta-\varphi)$ tends to $+\infty$, so $z\to+\infty$. By continuity, the image of $(-\alpha,\beta)$ is all of $\mathbb{R}$. Solving \eqref{eq:zi} for $\varphi$ gives the inverse
\begin{equation}\label{eq:zinv}
\varphi=\frac{\alpha\beta(e^{z}-1)}{\beta+\alpha e^{z}},
\end{equation}
whose denominator is positive for every $z\in\mathbb{R}$, so the inverse is smooth as well, and the componentwise map is a smooth diffeomorphism onto $\mathbb{R}^{m}$. At $\varphi=0$ the argument of the logarithm in \eqref{eq:zi} equals $\beta\alpha/(\alpha\beta)=1$, so $z=0$, and injectivity excludes any other zero. Outside $(-\alpha,\beta)$ the argument of the logarithm is zero, negative or undefined, so $z$ is finite exactly when $\varphi\in(-\alpha,\beta)$.
From \eqref{eq:errors} and \eqref{eq:abi}, $\alpha_{i}+\varphi_{i}=x_{i}-\underline{x}_{i}$ and $\beta_{i}-\varphi_{i}=\overline{x}_{i}-x_{i}$ for $i\geq2$, and the same holds for $i=1$ by \eqref{eq:ab1}, so the admissible interval of $\varphi_{i}$ is exactly the envelope of $x_{i}$. Dividing \eqref{eq:zinv} by $z$ gives \eqref{eq:kappa}. The function $(e^{z}-1)/z$ extends smoothly to $z=0$ with the value $1$, and it is positive because $e^{z}-1$ and $z$ share the same sign. The denominator $\beta+\alpha e^{z}$ is positive, so $\kappa_{i,l}$ is smooth and positive, and its value at $z=0$ is $\alpha\beta/(\alpha+\beta)$. 
\end{proof}

Note that, by \Cref{lem:coord}, the admissible interval of $\varphi_{i}$ coincides with the envelope of $x_{i}$. Hence, unlike BLF backstepping \cite{doi:10.1080/00207179.2011.631192,doi:10.1016/j.automatica.2015.10.034,doi:10.1109/TSMC.2016.2633007}, no band of fixed width around the virtual control is reserved, and no allocation of the envelope is needed. However, the coordinate $z_{i+1}$ is defined only while $\alpha_{i+1}\succ0$ and $\beta_{i+1}\succ0$, that is, while
\begin{equation}\label{eq:sep}
\underline{x}_{i+1}(t)<\eta_{i}(t)<\overline{x}_{i+1}(t),\qquad i=1,\ldots,n-1,
\end{equation}
which is the counterpart of \Cref{ass:ref} at block $i+1$, with the virtual control $\eta_{i}$ in place of the reference. Equivalently, $\nrm{(\eta_{i}-\check{x}_{i+1})/r_{i+1}}_{\infty}<1$. In other words, \eqref{eq:sep} requires the virtual control only to lie inside the envelope, whereas the feasibility condition of BLF backstepping requires it to keep the barrier width away from the envelope. Moreover, the ratio $\kappa_{i}$ in \eqref{eq:kappa} carries the cross term from block $i-1$ into the coordinate $z_{i}$.

We now use \emph{backstepping} based control to design the time-varying APC strategy. The design proceeds in $n+1$ steps, where the virtual controls $\eta_{1},\ldots,\eta_{n}$ are designed in Steps $1,\ldots,n$, and the command $u_{c}$ of the APIR is designed in Step $n+1$. In what follows, we select the weights $\mu_{i}>0$, $i=1,\ldots,n$, and define $R_{i}\coloneqq\diag(\kappa_{i,l})$ and the cross terms $\chi_{i}\coloneqq\mu_{i}G_{i}^{\top}Q_{i}z_{i}$.

\textit{Step 1}: One can obtain the time derivative of $\varphi_{1}$ using \eqref{eq:plant} and $x_{2}=\varphi_{2}+\eta_{1}$ as
\begin{equation}\label{eq:phi1}
\dot{\varphi}_{1}=\dot{x}_{1}-\dot{y}_{d}=f_{1}+G_{1}(\varphi_{2}+\eta_{1})-\dot{y}_{d}.
\end{equation}
By viewing $x_{2}$ as a virtual control input, let us consider a Lyapunov function candidate $\mathcal{W}_{1}=\tfrac12\mu_{1}\nrm{z_{1}}^{2}$, whose time derivative along the state trajectories results in $\dot{\mathcal{W}}_{1}=\mu_{1}z_{1}^{\top}\dot{z}_{1}$. If we design the virtual control $\eta_{1}$ as
\begin{equation}\label{eq:eta1}
\eta_{1}=G_{1}^{-1}\left[\dot{y}_{d}-f_{1}-Q_{1}^{-1}\left(\psi_{1}+k_{1}z_{1}\right)\right],\quad k_{1}>0,
\end{equation}
then substituting \eqref{eq:phi1} and \eqref{eq:eta1} into \eqref{eq:zdot} with $i=1$ yields the closed-loop dynamics, $\dot{z}_{1}=Q_{1}[f_{1}+G_{1}\varphi_{2}+G_{1}\eta_{1}-\dot{y}_{d}]+\psi_{1}=Q_{1}[G_{1}\varphi_{2}-Q_{1}^{-1}(\psi_{1}+k_{1}z_{1})]+\psi_{1}=-k_{1}z_{1}+Q_{1}G_{1}\varphi_{2}$, such that
the derivative of $\mathcal{W}_{1}$ becomes
\begin{align}
\dot{\mathcal{W}}_{1}&=-\mu_{1}k_{1}\nrm{z_{1}}^{2}+\mu_{1}z_{1}^{\top}Q_{1}G_{1}\varphi_{2}\nonumber\\
&=-\mu_{1}k_{1}\nrm{z_{1}}^{2}+\chi_{1}^{\top}\varphi_{2},\label{eq:W1}
\end{align}
where the last equality follows from $Q_{1}=Q_{1}^{\top}$ and the definition of $\chi_{1}$. Note that the cross term $\chi_{1}^{\top}\varphi_{2}$ in \eqref{eq:W1} is sign indefinite, and it is cancelled in the next step.

\textit{Step $i$ ($2\leq i\leq n$)}: In the $i$\textsuperscript{th} step, the error $\varphi_{i}=x_{i}-\eta_{i-1}$, on differentiating with respect to time and using \eqref{eq:plant}, yields
\begin{equation}\label{eq:phii}
\dot{\varphi}_{i}=\dot{x}_{i}-\dot{\eta}_{i-1}=f_{i}+G_{i}(\varphi_{i+1}+\eta_{i})-\dot{\eta}_{i-1},
\end{equation}
where $\varphi_{n+1}\coloneqq\varrho$, since the next block at $i=n$ is the realized input. Note that $\eta_{i-1}$ is a smooth function of $\bar{x}_{i-1}$ and $t$, so its derivative along \eqref{eq:plant} is available analytically as
\begin{equation}\label{eq:etadot}
\dot{\eta}_{i-1}=\sum_{j=1}^{i-1}\frac{\partial\eta_{i-1}}{\partial x_{j}}\left(f_{j}+G_{j}x_{j+1}\right)+\frac{\partial\eta_{i-1}}{\partial t},
\end{equation}
which enters both \eqref{eq:phii} and, through $\dot{\alpha}_{i}$ and $\dot{\beta}_{i}$, the function $\psi_{i}$. By considering $x_{i+1}$ a virtual control input to the $i$\textsuperscript{th} subsystem, we choose a Lyapunov function candidate as $\mathcal{W}_{i}=\mathcal{W}_{i-1}+\tfrac12\mu_{i}\nrm{z_{i}}^{2}$. Its time derivative contains the cross term $\chi_{i-1}^{\top}\varphi_{i}$ left by the preceding step, as in \eqref{eq:W1} for $i=2$. Since $\varphi_{i}=R_{i}z_{i}$ by \Cref{lem:coord}, this term can be written as $\chi_{i-1}^{\top}\varphi_{i}=\chi_{i-1}^{\top}R_{i}z_{i}$. The virtual control $\eta_{i}$ is therefore designed as
\begin{align}\label{eq:etai}
\eta_{i}={}&G_{i}^{-1}\left[\dot{\eta}_{i-1}-f_{i}-Q_{i}^{-1}\left(\psi_{i}+k_{i}z_{i}+\mu_{i}^{-1}R_{i}\chi_{i-1}\right)\right],
\end{align}
where $k_{i}>0$, which, on substituting \eqref{eq:phii} and \eqref{eq:etai} into \eqref{eq:zdot}, yields the closed-loop dynamics
\begin{align}
\dot{z}_{i}&=Q_{i}\left[f_{i}+G_{i}\varphi_{i+1}+G_{i}\eta_{i}-\dot{\eta}_{i-1}\right]+\psi_{i}\nonumber\\
&=Q_{i}G_{i}\varphi_{i+1}-\left(\psi_{i}+k_{i}z_{i}+\mu_{i}^{-1}R_{i}\chi_{i-1}\right)+\psi_{i}\nonumber\\
&=-k_{i}z_{i}-\mu_{i}^{-1}R_{i}\chi_{i-1}+Q_{i}G_{i}\varphi_{i+1}.\label{eq:zicl}
\end{align}
By differentiating $\mathcal{W}_{i}$ with respect to time along the state trajectories and using \eqref{eq:zicl}, one has
\begin{align}
\dot{\mathcal{W}}_{i}&=\dot{\mathcal{W}}_{i-1}+\mu_{i}z_{i}^{\top}\dot{z}_{i}=-\sum_{j=1}^{i-1}\mu_{j}k_{j}\nrm{z_{j}}^{2}+\chi_{i-1}^{\top}\varphi_{i}\nonumber\\
&-\mu_{i}k_{i}\nrm{z_{i}}^{2}-z_{i}^{\top}R_{i}\chi_{i-1}+\chi_{i}^{\top}\varphi_{i+1}\nonumber\\
&=-\sum_{j=1}^{i}\mu_{j}k_{j}\nrm{z_{j}}^{2}+\chi_{i}^{\top}\varphi_{i+1},\label{eq:Wi}
\end{align}
where the last equality follows from the relation $z_{i}^{\top}R_{i}\chi_{i-1}=(R_{i}z_{i})^{\top}\chi_{i-1}=\varphi_{i}^{\top}\chi_{i-1}$, since $R_{i}$ is diagonal. In other words, the virtual control \eqref{eq:etai} cancels the cross term of the preceding step and adds the damping $-\mu_{i}k_{i}\nrm{z_{i}}^{2}$. At $i=n$, \eqref{eq:etadot} expresses $\dot{\eta}_{n}$ as a function of $\bar{x}_{n}$ and $u$.

\textit{Step $n+1$}: In the last step, the realization error is defined as $\varrho=u-\eta_{n}$, which on differentiating with respect to time and using \eqref{eq:apir}, yields
\begin{equation}\label{eq:rhodot}
\dot{\varrho}=\mathcal{G}(u,t)u_{c}-\mathcal{F}(u)-\dot{\eta}_{n}.
\end{equation}
Unlike $\varphi_{i}$, the error $\varrho$ requires no barrier coordinate transformation, since the admissibility of $u$ is guaranteed by \Cref{thm:apir}. Let us consider the Lyapunov function candidate $\mathcal{W}=\mathcal{W}_{n}+\tfrac12\nrm{\varrho}^{2}$. If we design the command $u_{c}$ as
\begin{align}\label{eq:uc}
u_{c}&=\mathcal{G}(u,t)^{-1}\mathcal{N},\nonumber\\
\mathcal{N}&\coloneqq\mathcal{F}(u)+\dot{\eta}_{n}-\chi_{n}-k_{n+1}\varrho,
\end{align}
where $k_{n+1}>0$, then substituting \eqref{eq:uc} into \eqref{eq:rhodot} gives $\dot{\varrho}=\mathcal{N}-\mathcal{F}(u)-\dot{\eta}_{n}=-\chi_{n}-k_{n+1}\varrho$. Consequently, using \eqref{eq:Wi} with $i=n$ and $\varphi_{n+1}=\varrho$, the derivative of $\mathcal{W}$ along the state trajectories becomes
\begin{align}
\dot{\mathcal{W}}&=\dot{\mathcal{W}}_{n}+\varrho^{\top}\dot{\varrho}\nonumber\\
&=-\sum_{j=1}^{n}\mu_{j}k_{j}\nrm{z_{j}}^{2}+\chi_{n}^{\top}\varrho-\varrho^{\top}\chi_{n}-k_{n+1}\nrm{\varrho}^{2}\nonumber\\
&=-\sum_{j=1}^{n}\mu_{j}k_{j}\nrm{z_{j}}^{2}-k_{n+1}\nrm{\varrho}^{2}\leq-2\vartheta\mathcal{W},\label{eq:Wexact}
\end{align}
where $\vartheta\coloneqq\min_{1\leq j\leq n+1}k_{j}$, and the inequality follows from $\mu_{j}k_{j}\nrm{z_{j}}^{2}\geq\vartheta\mu_{j}\nrm{z_{j}}^{2}$ and $k_{n+1}\nrm{\varrho}^{2}\geq\vartheta\nrm{\varrho}^{2}$. The identity holds wherever the closed-loop vector field is defined.

Note that the weights $\mu_{i}$ do not affect the decay rate in \eqref{eq:Wexact}, but they scale the cross terms $\chi_{i}$. When the state and its virtual control lie near the middle of the envelope, one has $\alpha_{i,l}\approx\beta_{i,l}\approx r_{i,l}$, and \eqref{eq:zi} gives $z_{i,l}\approx2\varphi_{i,l}/r_{i,l}$. Thus, a wide envelope maps a given physical error into a small $z_{i}$, so that equal weights overstate the cross terms between blocks of unequal width. This motivates the choice $\mu_{i}=\bar{r}_{i}^{2}/4$, where $\bar{r}_{i}$ is a representative half-width of $\mathbb{X}_{i}$. With this choice, $\tfrac12\mu_{i}\nrm{z_{i}}^{2}\approx\tfrac12\nrm{\varphi_{i}}^{2}$, and hence $\mathcal{W}$ approximates half the sum of the squared physical errors of all blocks and the realization.

The term $\dot{\eta}_{n}$ in \eqref{eq:uc} depends on every state block, the realized input, the reference, and the envelopes, and its size grows with $n$. Since $\eta_{1}$ contains $\dot{y}_{d}$, $\dot{\alpha}_{1}$, and $\dot{\beta}_{1}$ and is differentiated $n$ times, the recursion uses $n+1$ derivatives of $y_{d}$ and of the output envelope. Similarly, the envelope of $x_{i}$, $i\geq2$, enters $\eta_{i}$ with its first derivative through $\psi_{i}$, and $\eta_{i}$ is differentiated $n+1-i$ times, so $n+2-i$ derivatives of that envelope are used, as required by \Cref{ass:ref,ass:env}.

\begin{figure}[!t]
\centering
\resizebox{\linewidth}{!}{%
\begin{tikzpicture}[
>=Stealth, font=\small, x=1mm, y=1mm,
sub/.style={draw=blue!55!black, fill=white, rounded corners=2pt, line width=0.6pt,
  minimum width=30mm, minimum height=10mm, align=center, inner sep=1.5pt},
apir/.style={draw=orange!75!black, fill=orange!12, rounded corners=2pt, line width=0.8pt,
   minimum width=22mm, minimum height=12mm, align=center, inner sep=2pt},
plant/.style={draw=green!45!black, fill=green!10, rounded corners=2pt, line width=0.8pt,
    minimum width=15mm, minimum height=12mm, align=center, inner sep=2pt},
setb/.style={draw=black!55, dashed, rounded corners=2pt, line width=0.6pt, fill=black!3,
   minimum width=24mm, minimum height=9mm, align=center, inner sep=2pt},
lab/.style={font=\small},
sig/.style={-{Stealth[length=2.2mm]}, line width=0.75pt},
setsig/.style={sig, black!60},
fbu/.style={sig, blue!70!black},
fbx/.style={sig, black!55, densely dashed}]
\draw[black!45, densely dotted, line width=0.9pt, rounded corners=4pt] (-3,12) rectangle (64,-35);
\node[font=\small\itshape, text=black!70, anchor=north east] at (63.2,-1) {APC law};
\draw[blue!55!black, line width=0.8pt, rounded corners=3pt, fill=blue!6] (0,9) rectangle (33,-32);
\node[font=\small\bfseries, text=blue!45!black, anchor=north west] at (1,8.6) {Synthesis};
\node[sub] (s1) at (16.5,-2) {Barrier coordinates $z_{i}$\\[-1pt]{\footnotesize \eqref{eq:errors}--\eqref{eq:zi}}};
\node[sub] (s2) at (16.5,-14.5) {Virtual controls $\eta_{i}$, $\dot{\eta}_{i}$\\[-1pt]{\footnotesize Steps $1,\ldots,n$}};
\node[sub] (s3) at (16.5,-27) {Command $u_{c}$\\[-1pt]{\footnotesize Step $n+1$, \eqref{eq:uc}}};
\draw[sig] (s1) -- (s2);
\draw[sig] (s2) -- (s3);
\node[apir] (apir) at (50,-27) {\textbf{APIR} \eqref{eq:apir}\\[1pt]{\footnotesize $\dot{u}=\mathcal{G}u_{c}-\mathcal{F}$}};
\node[plant] (pl) at (79,-27) {\textbf{Plant}\\[1pt]{\footnotesize \eqref{eq:plant}}};
\node[setb] (xset) at (16.5,21) {$\mathbb{X}_{i}(t)$\\[-1pt]{\footnotesize $\underline{x}_{i}(t),\ \overline{x}_{i}(t)$}};
\node[setb] (uset) at (50,21) {$\mathbb{U}(t)$\\[-1pt]{\footnotesize $u_{\min}(t),\ u_{\max}(t)$}};
\draw[setsig] (xset) -- (16.5,9);
\draw[setsig] (uset) -- (apir);
\fill[black!60] (50,-14.5) circle (1.3pt);
\draw[setsig] (50,-14.5) -- (s2.east);
\draw[sig] (-10,0) node[left,lab]{$y_{d}$} -- (s1.west |- 0,0);
\draw[sig] (s3.east) -- node[above,lab]{$u_{c}$} (apir.west);
\draw[sig] (apir.east) -- node[above,lab,pos=0.55]{$u$} (pl.west);
\coordinate (ub) at (66.5,-27);
\fill[blue!70!black] (ub) circle (1.3pt);
\draw[fbu] (ub) |- (24,-39) node[pos=0.72,below,lab,text=blue!70!black]{$u$} -- (24,-32.2);
\draw[fbx] (pl.east) -- (91,-27) |- (-6,-45) node[pos=0.75,below,lab,text=black!65]{$x=(x_{1},\ldots,x_{n})$} |- (s1.west |- 0,-4);
\end{tikzpicture}}
\caption{Time-varying APC architecture.}
\label{fig:arch}
\end{figure}
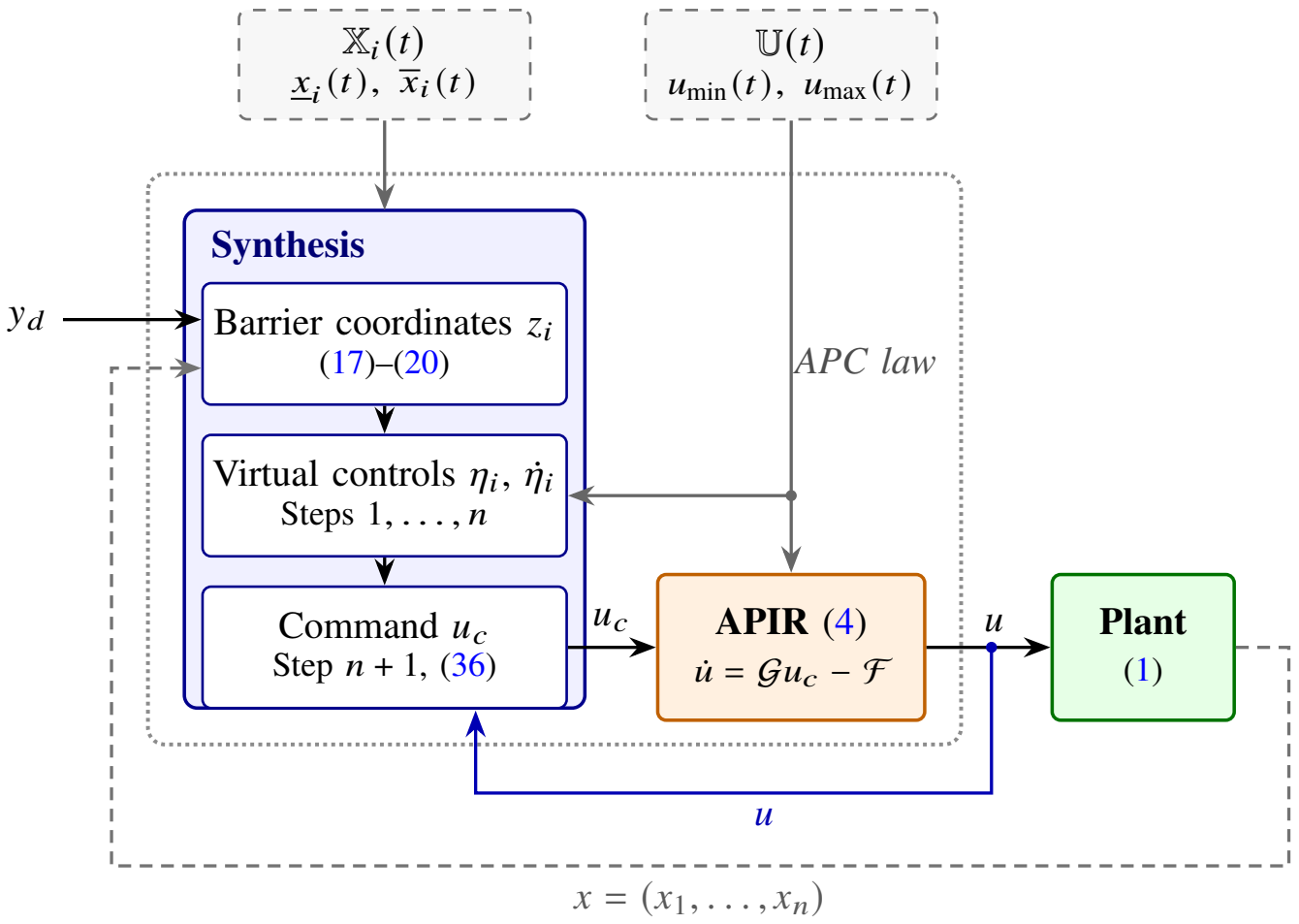

The resulting architecture of the time-varying APC law is shown in \Cref{fig:arch}. The synthesis block computes the barrier coordinate of each block from its state, virtual control, and envelope, generates the virtual controls in Steps $1$ to $n$, and forms the command $u_{c}$ in Step $n+1$. The APIR then converts $u_{c}$ into the admissible input $u$ applied to the plant. Note that the prescribed sets enter the controller as known signals, so the state envelopes and the actuator set are accounted for in the synthesis rather than enforced on its output.

\subsection{APC Compatibility and Closed-Loop Guarantees}\label{subsec:compat}
To close the dependence between the admissibility of the realization and the boundedness of the command, we introduce the sublevel set $\Omega_{c}\coloneqq\{\mathbf{X}\mid\mathcal{W}(\mathbf{X})\leq c\}$, $c>0$, where $\mathbf{X}=[z_{1}^{\top},\ldots,z_{n}^{\top},\varrho^{\top}]^{\top}$. Note that every point of $\Omega_{c}$, together with $t$, determines the plant state and the realized input recursively through $x_{1}=y_{d}+R_{1}z_{1}$, $x_{i+1}=\eta_{i}(\bar{x}_{i},t)+R_{i+1}z_{i+1}$, $i=1,\ldots,n-1$, and $u_{\mathbf{X}}=\eta_{n}(\bar{x}_{n},t)+\varrho$. This recursion is well defined as long as \eqref{eq:sep} holds, since $R_{i+1}$ depends on the widths \eqref{eq:abi}. Moreover, by \Cref{lem:coord}, every state obtained in this way lies in its envelope. For a fixed margin $\sigma_{\eta}\in(0,1)$, we call the level $c$ \emph{APC-compatible} if
\begin{equation}\label{eq:vcompat}
P_{c}\coloneqq\sup_{\mathbf{X}\in\Omega_{c},t\geq0} \max_{1\leq i\leq n-1}\nrm{\frac{\eta_{i}(\mathbf{X},t)-\check{x}_{i+1}(t)}{r_{i+1}(t)}}_{\infty}\leq1-\sigma_{\eta},
\end{equation}
and there exist a compact set $\mathbb{K}_{c}$, with $\mathbb{K}_{c}\subset\mathbb{U}(t)$ for all $t\geq0$, and a constant $\xi>0$ such that $u_{\mathbf{X}}\in\mathbb{K}_{c}$ for every $\mathbf{X}\in\Omega_{c}$ and $t\geq0$, and
\begin{equation}\label{eq:compat}
Q_{c}(\xi)\coloneqq\sup_{\mathbf{X}\in\Omega_{c},t\geq0}\nrm{\mathcal{G}(u_{\mathbf{X}},t)^{-1}\mathcal{N}(\mathbf{X},u_{\mathbf{X}},t)}_{\infty}\leq\xi .
\end{equation}
Note that \eqref{eq:vcompat} is the separation condition \eqref{eq:sep} with a margin, evaluated over the sublevel set, whereas \eqref{eq:compat} is the compatibility condition of \cite{doi:10.48550/arXiv.2608.15375} for moving sets. In other words, the two conditions couple the initial-error level $c$ and the reference demand to the state envelopes and to the finite actuator authority. A more conservative but easier sufficient test for \eqref{eq:compat} is $M_{c}\leq\xi\underline{\mathcal{G}}$, where $M_{c}$ is the supremum of $\nrm{\mathcal{N}}_{\infty}$ over the same set and $\underline{\mathcal{G}}$ is given by \eqref{eq:gainbound}.

\begin{remark}\label{rem:compat}
Note that \eqref{eq:vcompat} cannot be removed if exact tracking is required, since maintaining $x_{1}\equiv y_{d}$ requires $x_{2}=G_{1}^{-1}(\dot{y}_{d}-f_{1})$, which may leave $\mathbb{X}_{2}(t)$, and the same holds at every block. Unlike the feasibility condition of BLF backstepping \cite{doi:10.1080/00207179.2011.631192,doi:10.1016/j.automatica.2015.10.034,doi:10.1109/TSMC.2016.2633007}, it carries no barrier width, and by \Cref{lem:coord} its violation does not violate the state constraints. The design in \cite{doi:10.1109/TAC.2018.2845707} avoids this condition at the cost of uniform ultimate boundedness. 
Condition \eqref{eq:compat} cannot be removed either, since $\mathcal{G}u_{c}=\mathcal{N}$ identically, so the command cancels the inward vector field on which \Cref{thm:apir} relies. Moreover, \eqref{eq:compat} becomes more restrictive as the actuator set contracts faster, through $\varepsilon_{0}$, and as the gains $k_{i}$ increase, through $\dot{\eta}_{n}$ in $\mathcal{N}$.
\end{remark}

\begin{proposition}[Local nonvacuity of APC compatibility]\label{prop:nonvacuity}
Consider the plant \eqref{eq:plant} with the APIR \eqref{eq:apir}, the virtual controls \eqref{eq:eta1} and \eqref{eq:etai}, and the command \eqref{eq:uc}, under \Cref{ass:model,ass:ref,ass:env,ass:actuator}. Suppose that $u_{\max}$ and $u_{\min}$ are bounded, and that there exists $\underline{r}>0$ such that $r_{i}(t)\succeq\underline{r}\mathbf{1}$, $i=1,\ldots,n$, for all $t\geq0$. Let $\eta_{i}^{*}(t)\coloneqq\eta_{i}(\mathbf{0},t)$, $i=1,\ldots,n$, denote the virtual controls evaluated at $\mathbf{X}=\mathbf{0}$. Moreover, suppose that there exist $\sigma^{*}\in(\sigma_{\eta},1)$, a compact set $\mathbb{K}\subset\mathbb{R}^{m}$, and constants $\delta>0$ and $\xi_{0}>0$ such that the following conditions hold for all $t\geq0$. First, $\nrm{(\eta_{i}^{*}(t)-\check{x}_{i+1}(t))/r_{i+1}(t)}_{\infty}\leq1-\sigma^{*}$, $i=1,\ldots,n-1$. Second, $\eta_{n}^{*}(t)\in\mathbb{K}$, and every $u$ with $\operatorname{dist}_{\infty}(u,\mathbb{K})\leq2\delta$ lies in $\mathbb{U}(t)$. Third, the zero-error command satisfies
\begin{equation}\label{eq:zerocompat}
Q_{0}\coloneqq\sup_{t\geq0}\nrm{\mathcal{G}(\eta_{n}^{*}(t),t)^{-1}\mathcal{N}(\mathbf{0},\eta_{n}^{*}(t),t)}_{\infty}<\xi_{0}.
\end{equation}
Then there exists $c^{*}>0$ such that every level $c\in(0,c^{*}]$ is APC-compatible with $\xi=\xi_{0}$ and $\mathbb{K}_{c}\coloneqq\{u\mid\operatorname{dist}_{\infty}(u,\mathbb{K})\leq\delta\}$.
\end{proposition}

\begin{proof}
Let $w(t)$ denote the vector formed by $y_{d}$, the state envelopes, the actuator bounds, and the derivatives of these signals that enter \eqref{eq:etadot} and \eqref{eq:uc}. By \Cref{ass:ref,ass:env} and the boundedness of $u_{\max}$ and $u_{\min}$, $w(t)$ remains in a bounded set, whose closure we denote by $\mathbb{W}$. By construction, each $\eta_{i}$, the input $u_{\mathbf{X}}$ determined by $\mathbf{X}$, and $\mathcal{N}$ are continuous functions of $(\mathbf{X},w)$ on the open set $\mathcal{O}$ on which every width in \eqref{eq:ab1} and \eqref{eq:abi} is positive.

The point $\mathbf{X}=\mathbf{0}$ lies uniformly inside $\mathcal{O}$. Indeed, at $\mathbf{X}=\mathbf{0}$, every $z_{i}$ vanishes, so that $x_{1}=y_{d}$ and $x_{i+1}=\eta_{i}^{*}$. Hence, the widths of the first block are at least $\varepsilon_{y}$ by \Cref{ass:ref}. Moreover, since $\alpha_{i+1}=r_{i+1}+(\eta_{i}^{*}-\check{x}_{i+1})$ and $\beta_{i+1}=r_{i+1}-(\eta_{i}^{*}-\check{x}_{i+1})$ by \eqref{eq:abi}, the first condition gives $\alpha_{i+1}\succeq\sigma^{*}r_{i+1}\succeq\sigma^{*}\underline{r}\mathbf{1}$ and, similarly, $\beta_{i+1}\succeq\sigma^{*}\underline{r}\mathbf{1}$. Consequently, there exists $\rho>0$ such that $\{\mathbf{X}\mid\nrm{\mathbf{X}}\leq\rho\}\times\mathbb{W}\subset\mathcal{O}$, and the functions above are uniformly continuous on this compact set.

Uniform continuity yields $\rho_{1}\in(0,\rho]$ such that, for all $\nrm{\mathbf{X}}\leq\rho_{1}$ and all $t\geq0$,
\begin{align}
&\nrm{\frac{\eta_{i}(\mathbf{X},t)-\eta_{i}^{*}(t)}{r_{i+1}(t)}}_{\infty}\leq\sigma^{*}-\sigma_{\eta},\quad i=1,\ldots,n-1,\label{eq:nva}\\
&\nrm{u_{\mathbf{X}}-\eta_{n}^{*}(t)}_{\infty}\leq\delta,\label{eq:nvb}\\
&\nrm{\mathcal{G}(u_{\mathbf{X}},t)^{-1}\mathcal{N}(\mathbf{X},u_{\mathbf{X}},t)-\mathcal{G}(\eta_{n}^{*},t)^{-1}\mathcal{N}(\mathbf{0},\eta_{n}^{*},t)}_{\infty}\nonumber\\
&\quad\leq\xi_{0}-Q_{0}.\label{eq:nvc}
\end{align}
In \eqref{eq:nva}, the division by $r_{i+1}\succeq\underline{r}\mathbf{1}$ preserves uniform continuity. For \eqref{eq:nvc}, note that $u_{\mathbf{X}}$ and $\eta_{n}^{*}$ lie in $\mathbb{K}_{c}$ by \eqref{eq:nvb} and the second condition, and every point of $\mathbb{K}_{c}$ lies at a distance of at least $\delta$ from the boundary of $\mathbb{U}(t)$. Since $u_{\max}$ and $u_{\min}$ are bounded, the diagonal entries of $\mathcal{G}$ in \eqref{eq:GF} are bounded below by a positive constant on $\mathbb{K}_{c}$, so that $\mathcal{G}^{-1}$ is uniformly continuous there.

Finally, since $\mathcal{W}\geq\tfrac12\underline{\mu}\nrm{\mathbf{X}}^{2}$ with $\underline{\mu}\coloneqq\min\{\mu_{1},\ldots,\mu_{n},1\}$, the choice $c^{*}\coloneqq\underline{\mu}\rho_{1}^{2}/2$ gives $\nrm{\mathbf{X}}\leq\rho_{1}$ on $\Omega_{c}$ for every $c\in(0,c^{*}]$. Hence, by the triangle inequality, the first condition and \eqref{eq:nva} give $P_{c}\leq(1-\sigma^{*})+(\sigma^{*}-\sigma_{\eta})=1-\sigma_{\eta}$, which is \eqref{eq:vcompat}. Moreover, \eqref{eq:nvb} gives $u_{\mathbf{X}}\in\mathbb{K}_{c}$, and \eqref{eq:nvc} together with \eqref{eq:zerocompat} gives $Q_{c}(\xi_{0})\leq Q_{0}+(\xi_{0}-Q_{0})=\xi_{0}$, which is \eqref{eq:compat}. Therefore, every level $c\in(0,c^{*}]$ is APC-compatible.
\end{proof}

Note that the hypotheses of \Cref{prop:nonvacuity} involve only the zero-error closed loop, that is, the reference, the envelopes, and the actuator bounds, so that they can be checked before the controller is run.

We first establish that the state constraints hold without any compatibility condition.

\begin{proposition}[State constraints on the interval of existence]\label{prop:exist}
Consider the plant \eqref{eq:plant} with the APIR \eqref{eq:apir}, the virtual controls \eqref{eq:eta1} and \eqref{eq:etai}, and the command \eqref{eq:uc}, under \Cref{ass:model,ass:ref,ass:env,ass:actuator}. Let $x_{i}(0)\in\mathbb{X}_{i}(0)$, $i=1,\ldots,n$, and $u(0)\in\mathbb{U}(0)$, and let \eqref{eq:sep} hold at $t=0$. Then $x_{i}(t)\in\mathbb{X}_{i}(t)$, $i=1,\ldots,n$, for all $t\in[0,T_{\max})$, where $[0,T_{\max})$ is the maximal interval of existence of the closed-loop solution.
\end{proposition}

\begin{proof}
The closed-loop vector field is defined on the open set $\mathcal{D}$ on which every width $\alpha_{i}$, $\beta_{i}$ is positive, every coordinate $z_{i}$ is finite, and $\mathcal{G}(u,t)$ is invertible. By \Cref{ass:ref}, \eqref{eq:sep} at $t=0$, and $u(0)\in\mathbb{U}(0)$, the initial condition lies in $\mathcal{D}$. Hence, the solution remains in $\mathcal{D}$ on its maximal interval of existence $[0,T_{\max})$, so that every $z_{i}(t)$ is finite. By \Cref{lem:coord}, a finite $z_{i}$ is equivalent to $\underline{x}_{i}(t)<x_{i}(t)<\overline{x}_{i}(t)$, which yields $x_{i}(t)\in\mathbb{X}_{i}(t)$, $i=1,\ldots,n$, for all $t\in[0,T_{\max})$.
\end{proof}

Note that, by \Cref{prop:exist}, no state leaves its envelope while the solution exists. In other words, if compatibility fails, then either the input leaves $\mathbb{U}(t)$ or the solution ceases to exist in finite time, that is, $T_{\max}<\infty$.

\begin{theorem}[Exponential admissibility-preserving tracking]\label{thm:main}
Consider the plant \eqref{eq:plant} with the APIR \eqref{eq:apir}, the virtual controls \eqref{eq:eta1} and \eqref{eq:etai}, and the command \eqref{eq:uc}, under \Cref{ass:model,ass:ref,ass:env,ass:actuator}. Let $x_{i}(0)\in\mathbb{X}_{i}(0)$, $i=1,\ldots,n$, and $u(0)\in\mathbb{U}(0)$, and let \eqref{eq:sep} hold at $t=0$. Moreover, let the level $c$ be APC-compatible with $\mathcal{W}(0)<c$, and define $\bar{\omega}_{1}\coloneqq\sup_{t\geq0}\nrm{\overline{x}_{1}(t)-\underline{x}_{1}(t)}_{\infty}$. Then the closed-loop solution exists for all $t\geq0$, $\Omega_{c}$ is forward invariant, \eqref{eq:Wexact} holds for all $t\geq0$, and the tracking error satisfies
\begin{equation}\label{eq:track}
\nrm{\varphi_{1}(t)}_{\infty}\leq\frac{\bar{\omega}_{1}}{4}\sqrt{\frac{2\mathcal{W}(0)}{\mu_{1}}}e^{-\vartheta t}.
\end{equation}
Moreover, for all $t\geq0$, $x_{i}(t)\in\mathbb{X}_{i}(t)$, $i=1,\ldots,n$, the realized input satisfies \eqref{eq:confine}, and hence $u(t)\in\mathbb{U}(t)$, and all closed-loop signals are bounded.
\end{theorem}

\begin{proof}
Let $[0,T)$ denote the maximal interval on which the solution exists with $\mathbf{X}(t)\in\Omega_{c}$ and $u(t)\in\mathbb{U}(t)$. On this interval, \eqref{eq:Wexact} gives $\mathcal{W}(t)\leq\mathcal{W}(0)e^{-2\vartheta t}<c$, so that $\mathbf{X}(t)$ remains in the interior of $\Omega_{c}$. Hence, by \eqref{eq:vcompat}, every width $\alpha_{i+1}$ and $\beta_{i+1}$ is at least $\sigma_{\eta}r_{i+1}$, and every coordinate $z_{i}$ is defined. Since each $z_{i}$ is bounded and each width is bounded below, \Cref{prop:exist} implies that every state remains at a uniform distance from its envelope. Moreover, compatibility gives $\nrm{u_{c}(t)}_{\infty}\leq Q_{c}(\xi)\leq\xi$ on $[0,T)$. Therefore, \Cref{thm:apir}, applied on this interval, yields \eqref{eq:confine} and \eqref{eq:gainbound}, so that $u$ remains at a uniform distance from the boundary of $\mathbb{U}(t)$. Consequently, the closed-loop state is confined to a compact subset of the open domain of the vector field, and the extension theorem \cite[Th.~3.3]{khalil2002nonlinear} gives $T=\infty$.

To obtain \eqref{eq:track}, note from \eqref{eq:q} that $q_{1,l}\geq4/(\alpha_{1,l}+\beta_{1,l})$ on $(-\alpha_{1,l},\beta_{1,l})$, with equality at the midpoint. Integrating $q_{1,l}$ from $0$ to $\varphi_{1,l}$ gives $\abs{z_{1,l}}\geq4\abs{\varphi_{1,l}}/(\alpha_{1,l}+\beta_{1,l})$. Since $\alpha_{1,l}+\beta_{1,l}=\overline{x}_{1,l}-\underline{x}_{1,l}\leq\bar{\omega}_{1}$ and $\mu_{1}\nrm{z_{1}}^{2}\leq2\mathcal{W}\leq2\mathcal{W}(0)e^{-2\vartheta t}$, \eqref{eq:track} follows. Finally, the remaining closed-loop signals are bounded by the smoothness of $f_{i}$, $G_{i}$, and the virtual controls on the compact set obtained above.
\end{proof}

\begin{corollary}[Explicit state margins and input rate]\label{cor:margins}
Consider the plant \eqref{eq:plant} with the APIR \eqref{eq:apir}, the virtual controls \eqref{eq:eta1} and \eqref{eq:etai}, and the command \eqref{eq:uc}, under \Cref{ass:model,ass:ref,ass:env,ass:actuator}. Suppose that the initial conditions satisfy $x_{i}(0)\in\mathbb{X}_{i}(0)$, $i=1,\ldots,n$, $u(0)\in\mathbb{U}(0)$, and \eqref{eq:sep} at $t=0$, and that the level $c$ is APC-compatible with $\mathcal{W}(0)<c$. For $i=1,\ldots,n$, define $\bar{z}_{i}(t)\coloneqq\sqrt{2\mathcal{W}(0)/\mu_{i}}\,e^{-\vartheta t}$, and let the relative margin of the $l$th component of $x_{i}$ be $m_{i,l}\coloneqq\min\{x_{i,l}-\underline{x}_{i,l},\,\overline{x}_{i,l}-x_{i,l}\}/(\overline{x}_{i,l}-\underline{x}_{i,l})$. Moreover, let $\bar{\omega}_{1}\coloneqq\sup_{t\geq0}\nrm{\overline{x}_{1}(t)-\underline{x}_{1}(t)}_{\infty}$. Then, for every component $l$ and all $t\geq0$,
\begin{align}
m_{1,l}(t)&\geq\frac{\varepsilon_{y}}{\varepsilon_{y}+(\bar{\omega}_{1}-\varepsilon_{y})e^{\bar{z}_{1}(t)}},\label{eq:margin1}\\
m_{i,l}(t)&\geq\frac{\sigma_{\eta}}{\sigma_{\eta}+(2-\sigma_{\eta})e^{\bar{z}_{i}(t)}},\quad i\geq2.\label{eq:margini}
\end{align}
where $\varepsilon_{y}$ and $\sigma_{\eta}$ are given in \Cref{ass:ref} and \eqref{eq:vcompat}, respectively. If, in addition, $u_{\max}$ and $u_{\min}$ are bounded, then the realized input satisfies
\begin{equation}\label{eq:udot}
\nrm{\dot{u}(t)}_{\infty}\leq p_{1}\left(\xi+p_{2}\bar{u}\right),
\end{equation}
where $\xi$ is the command bound in \eqref{eq:compat}, $\bar{u}\coloneqq\max\{\bar{s}\sup_{t,l}u_{\max,l}(t),\,\bar{\zeta}\sup_{t,l}\abs{u_{\min,l}(t)}\}$, and $\bar{s}$ and $\bar{\zeta}$ are given in \eqref{eq:zbar}.
\end{corollary}

\begin{proof}
From the definition of $\mathcal{W}$ and \Cref{thm:main}, for every $i=1,\ldots,n$, $\tfrac12\mu_{i}\nrm{z_{i}(t)}^{2}\leq\mathcal{W}(t)\leq\mathcal{W}(0)e^{-2\vartheta t},
$, where the second inequality follows from \eqref{eq:Wexact} by the comparison lemma. Hence, $\abs{z_{i,l}(t)}\leq\nrm{z_{i}(t)}\leq\bar{z}_{i}(t)$ for every component $l$.

Next, from \eqref{eq:errors}, \eqref{eq:ab1}, and \eqref{eq:abi}, one has $x_{i,l}-\underline{x}_{i,l}=\alpha_{i,l}+\varphi_{i,l}$ and $\overline{x}_{i,l}-x_{i,l}=\beta_{i,l}-\varphi_{i,l}$. Substituting the inverse \eqref{eq:zinv} gives
\begin{align}
x_{i,l}-\underline{x}_{i,l}&=\frac{\alpha_{i,l}(\alpha_{i,l}+\beta_{i,l})e^{z_{i,l}}}{\beta_{i,l}+\alpha_{i,l}e^{z_{i,l}}},\label{eq:distlo}\\
\overline{x}_{i,l}-x_{i,l}&=\frac{\beta_{i,l}(\alpha_{i,l}+\beta_{i,l})}{\beta_{i,l}+\alpha_{i,l}e^{z_{i,l}}},\label{eq:disthi}
\end{align}
where $\alpha_{i,l}+\beta_{i,l}=\overline{x}_{i,l}-\underline{x}_{i,l}$. Dividing \eqref{eq:distlo} and \eqref{eq:disthi} by $\alpha_{i,l}+\beta_{i,l}$, and suppressing the indices, the relative margin becomes $m=\min\left\{\frac{\alpha}{\alpha+\beta e^{-z}},\ \frac{\beta}{\beta+\alpha e^{z}}\right\}.$ The first term increases and the second decreases with $z$. Since $-\bar{z}_{i}\leq z\leq\bar{z}_{i}$, evaluating the first term at $z=-\bar{z}_{i}$ and the second at $z=\bar{z}_{i}$ yields
\begin{equation}\label{eq:mlow}
m\geq\min\left\{\frac{\alpha}{\alpha+\beta e^{\bar{z}_{i}}},\ \frac{\beta}{\beta+\alpha e^{\bar{z}_{i}}}\right\}.
\end{equation}
Writing $\beta=\omega-\alpha$ with $\omega\coloneqq\alpha+\beta$, the first term in \eqref{eq:mlow} equals $\alpha/(\alpha+(\omega-\alpha)e^{\bar{z}_{i}})$, which increases with $\alpha$ and decreases with $\omega$, since $e^{\bar{z}_{i}}\geq1$. The same holds for the second term with $\alpha$ and $\beta$ interchanged. Therefore, if $\alpha\geq\underline{w}$, $\beta\geq\underline{w}$, and $\omega\leq\bar{\omega}$, then
\begin{equation}\label{eq:mgen}
m\geq\frac{\underline{w}}{\underline{w}+(\bar{\omega}-\underline{w})e^{\bar{z}_{i}}}.
\end{equation}
For $i=1$, \Cref{ass:ref} and \eqref{eq:ab1} give $\alpha_{1,l}\geq\varepsilon_{y}$ and $\beta_{1,l}\geq\varepsilon_{y}$, and $\alpha_{1,l}+\beta_{1,l}=\overline{x}_{1,l}-\underline{x}_{1,l}\leq\bar{\omega}_{1}$. Hence, \eqref{eq:mgen} with $\underline{w}=\varepsilon_{y}$ and $\bar{\omega}=\bar{\omega}_{1}$ gives \eqref{eq:margin1}. For $i\geq2$, \eqref{eq:abi} gives $\alpha_{i,l}=r_{i,l}+(\eta_{i-1,l}-\check{x}_{i,l})$ and $\beta_{i,l}=r_{i,l}-(\eta_{i-1,l}-\check{x}_{i,l})$. Since $\Omega_{c}$ is forward invariant by \Cref{thm:main}, \eqref{eq:vcompat} gives $\abs{\eta_{i-1,l}-\check{x}_{i,l}}\leq(1-\sigma_{\eta})r_{i,l}$, so that $\alpha_{i,l}\geq\sigma_{\eta}r_{i,l}$, $\beta_{i,l}\geq\sigma_{\eta}r_{i,l}$, and $\alpha_{i,l}+\beta_{i,l}=2r_{i,l}$. Hence, \eqref{eq:mgen} with $\underline{w}=\sigma_{\eta}r_{i,l}$ and $\bar{\omega}=2r_{i,l}$ gives \eqref{eq:margini}.

Finally, by \eqref{eq:apir}, $\dot{u}_{l}=p_{1}\mathcal{S}_{l}u_{c,l}-p_{1}p_{2}u_{l}$, where $0<\mathcal{S}_{l}\leq1$ along the solution by \eqref{eq:confine}. Hence, $\abs{\dot{u}_{l}}\leq p_{1}\abs{u_{c,l}}+p_{1}p_{2}\abs{u_{l}}$. Since $\Omega_{c}$ is forward invariant, \eqref{eq:compat} gives $\nrm{u_{c}}_{\infty}\leq\xi$, and \eqref{eq:confine} gives $\abs{u_{l}}\leq\bar{u}$, which yields \eqref{eq:udot}.
\end{proof}

Note that, as $t\to\infty$, the bounds in \eqref{eq:margin1} and \eqref{eq:margini} approach $\varepsilon_{y}/\bar{\omega}_{1}$ and $\sigma_{\eta}/2$, respectively, since $\bar{z}_{i}(t)\to0$.

\begin{remark}\label{rem:exact}
Note that \Cref{thm:main} extends the result of \cite{doi:10.48550/arXiv.2608.15375} to full state constraints and moving sets. Since \eqref{eq:Wexact} is exact, the tracking error converges to zero for a time-varying reference without any boundary-layer residual. However, the design requires the analytic expansion of $\dot{\eta}_{n}$, which grows with $n$, and smoothness of the reference and of every envelope. Relaxing \eqref{eq:vcompat} by filtering the recursion, as in \cite{doi:10.1109/TAC.2018.2845707}, would yield uniform ultimate boundedness instead of exponential convergence.
\end{remark}

\begin{corollary}[Transfer of input safety under realization error]\label{cor:tight}
Let the input applied by the physical actuator be  {$u_{a}=u+d_{u}$}, where the realization error satisfies $\nrm{d_{u}(t)}_{\infty}\leq\bar{d}_{u}$ for all $t\geq0$. Let the APIR \eqref{eq:apir} be designed on the tightened interval $\widetilde{\mathbb{U}}(t)\coloneqq(u_{\min}(t)+\bar{d}_{u}\mathbf{1},u_{\max}(t)-\bar{d}_{u}\mathbf{1})$, with a piecewise-continuous command satisfying $\nrm{u_{c}(t)}_{\infty}\leq\xi$ for all $t\geq0$. Moreover, let the tightened bounds satisfy \Cref{ass:actuator}, that is, $u_{\max,l}-\bar{d}_{u}\geq\underline{u}_{\max}>0$, $u_{\min,l}+\bar{d}_{u}\leq-\underline{u}_{\min}<0$, and
\begin{align}\label{eq:tightrate}
\frac{\dot{u}_{\max,l}}{u_{\max,l}-\bar{d}_{u}}&\geq-p_{1}p_{2}+\varepsilon_{0},\nonumber\\
\frac{\dot{u}_{\min,l}}{u_{\min,l}+\bar{d}_{u}}&\geq-p_{1}p_{2}+\varepsilon_{0},
\end{align}
for every component $l$ and all $t\geq0$. Then $u_{\mathrm{phys}}(t)\in\mathbb{U}(t)$ for all $t\geq0$.
\end{corollary}

\begin{proof}
Applying \Cref{thm:apir} to the tightened bounds yields $u(t)\in\widetilde{\mathbb{U}}(t)$ for all $t\geq0$. Since $\nrm{d_{u}(t)}_{\infty}\leq\bar{d}_{u}$, it follows that {$u_{a}(t)=u(t)+d_{u}(t)\in\mathbb{U}(t)$}.
\end{proof}

Note that \eqref{eq:tightrate} does not follow from \eqref{eq:rate2}. Since tightening divides the rate of a contracting boundary by a smaller positive number, bounds that satisfy the rate condition on $\mathbb{U}(t)$ may violate it on $\widetilde{\mathbb{U}}(t)$. Hence, the margin $\varepsilon_{0}$ must be large enough to absorb the factor $u_{\max,l}/(u_{\max,l}-\bar{d}_{u})$, so that tightening reduces both the available authority and the rate margin. 
\section{Numerical Simulations}\label{sec:sim}
In this section, we demonstrate the performance of the proposed time-varying APC strategy through numerical simulations. We first consider a scenario in which the state envelopes translate over time while the actuator set oscillates and varies in width. We then investigate the effect of the control gains on the performance of the proposed strategy. Thereafter, we consider a scenario in which the actuator set contracts at different rates, including rates reaching $95\%$ of the admissible rate $p_{1}p_{2}$ specified in \Cref{ass:actuator}. Finally, we investigate the sensitivity of the design to the choice of weights $\mu_i$ under both oscillating and contracting actuator sets. In all cases, we consider the plant \eqref{eq:plant} with $n=2$ and $m=2$, where
\begin{align*}
f_{1}&=\begin{bmatrix}0.5\sin x_{1,1}\\ -0.3x_{1,2}+0.2x_{1,1}^{2}\end{bmatrix},\,
G_{1}=\begin{bmatrix}1.2+0.1x_{1,1}^{2} & 0.2\\ -0.1 & 1.0\end{bmatrix},\\
f_{2}&=\begin{bmatrix}0.4x_{2,1}x_{1,1}\\ -0.5x_{2,2}+0.3\sin x_{1,2}\end{bmatrix},\,
G_{2}=\begin{bmatrix}1.0 & 0.3\\ 0.2 & 1.1\end{bmatrix}.
\end{align*}
The reference trajectory is chosen as $y_{d}=[0.5\sin0.5t,~0.4\cos0.4t]^{\top}$. The state envelopes are defined by $\underline{x}_{1}=(-1+0.3\sin0.6t)\mathbf{1}$, $\overline{x}_{1}=(1+0.3\sin0.6t)\mathbf{1}$, $\underline{x}_{2}=(-2.5+0.4\sin0.35t)\mathbf{1}$, and $\overline{x}_{2}=(2.5+0.4\sin0.35t)\mathbf{1}$. The actuator bounds are given by $u_{\max}=(3+\sin0.3t)\mathbf{1}$ and $u_{\min}=-(2.5+0.8\cos0.25t)\mathbf{1}$. Thus, both state envelopes translate over time while maintaining constant widths, whereas the actuator set has simultaneous translation and width variation.

The APIR parameters are selected as $p_{1}=5$, $p_{2}=0.5$, and $\gamma=2$, which yield $p_{1}p_{2}=2.5$. Since $\sup_{t}\abs{\dot{u}_{\max,l}/u_{\max,l}}\leq 0.15$ and $\sup_{t}\abs{\dot{u}_{\min,l}/u_{\min,l}}\leq0.118$, \Cref{ass:actuator} holds. The controller gains are chosen as $k_{1}=k_{2}=2$ and $k_{3}=4$, such that $\vartheta=2$, and $\sigma_{\eta}=0.05$. The weights are selected as $\mu_{i}=\bar{r}_{i}^{2}/4$, where $\bar{r}_{1}=1$ and $\bar{r}_{2}=2.5$ denote the half-widths of the respective state envelopes. This implies $\mu_{1}=0.25$ and $\mu_{2}=1.5625$. The initial conditions are chosen as $x_{1}(0)=[0.35,-0.25]^{\top}$, $x_{2}(0)=[-0.501,1.238]^{\top}$, and $u(0)=[0.436,-0.680]^{\top}$. To quantify constraint satisfaction, we define the relative margin of a component $v_{l}$ with respect to its admissible interval as
\begin{equation*}
\mathcal{M}_{v,l}:=\inf_{t\geq0}\frac{\min\{v_{l}(t)-\underline{v}_{l}(t),\,\overline{v}_{l}(t)-v_{l}(t)\}}{\overline{v}_{l}(t)-\underline{v}_{l}(t)},
\end{equation*}
and, for a vector signal, take the minimum margin among its components.  Note that a positive margin indicates that the signal remains strictly inside its admissible set throughout the interval.

\begin{figure*}[!ht]
\centering
\begin{subfigure}{0.32\linewidth}
\centering
\includegraphics[width=\linewidth]{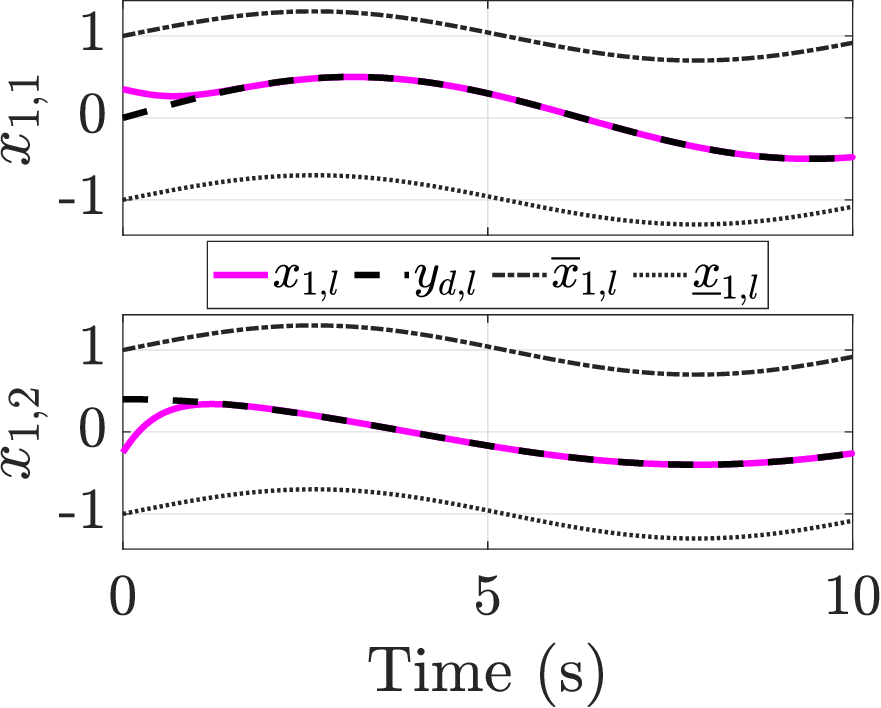}
\caption{First state block.}
\label{fig:states_x1}
\end{subfigure}%
\hfill
\begin{subfigure}{0.32\linewidth}
\centering
\includegraphics[width=\linewidth]{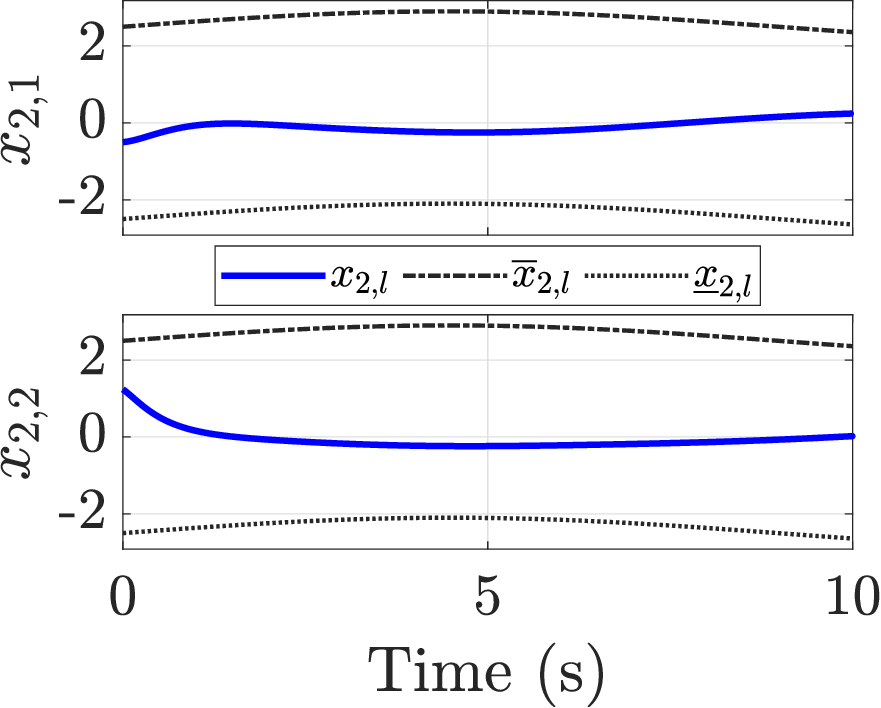}
\caption{Second state block.}
\label{fig:states_x2}
\end{subfigure}%
\hfill
\begin{subfigure}{0.32\linewidth}
\centering
\includegraphics[width=\linewidth]{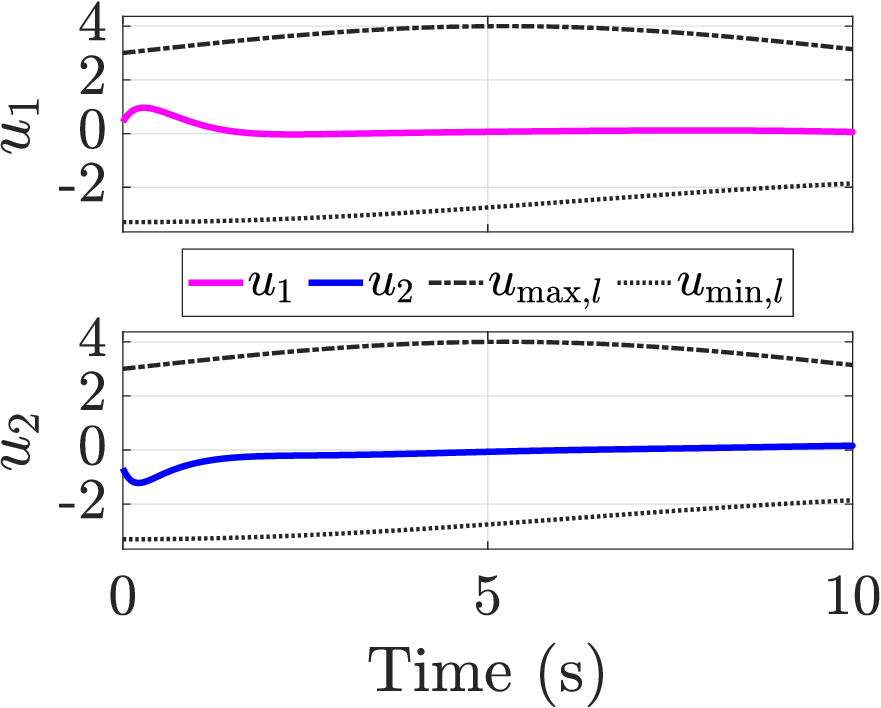}
\caption{Control inputs.}
\label{fig:states_u}
\end{subfigure}\\
\begin{subfigure}{0.32\linewidth}
\centering
\includegraphics[width=\linewidth]{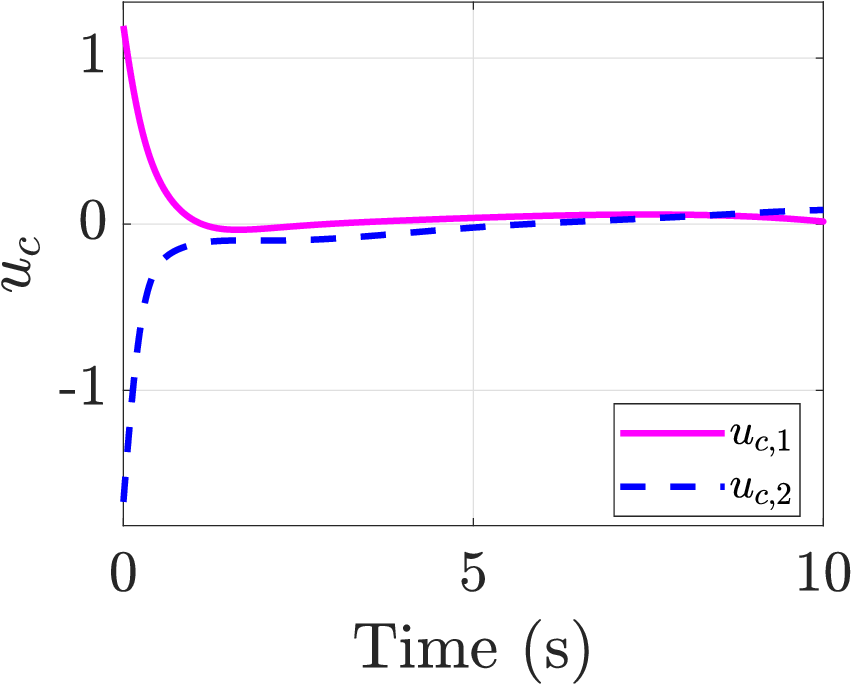}
\caption{Commanded inputs.}
\label{fig:states_uc}
\end{subfigure}%
\hfill
\begin{subfigure}{0.32\linewidth}
\centering
\includegraphics[width=\linewidth]{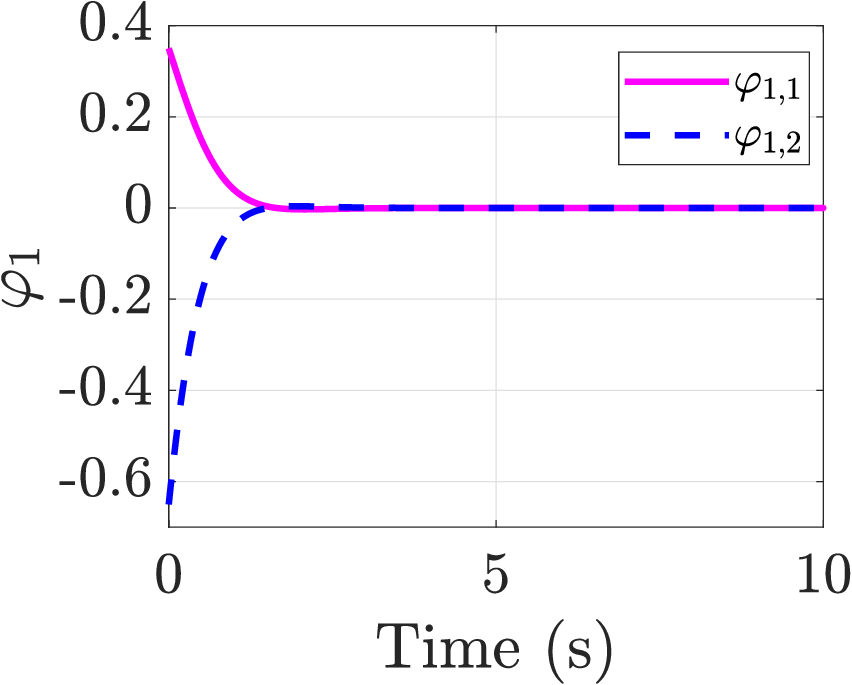}
\caption{Tracking errors.}
\label{fig:lyap_err}
\end{subfigure}%
\hfill
\begin{subfigure}{0.32\linewidth}
\centering
\includegraphics[width=\linewidth]{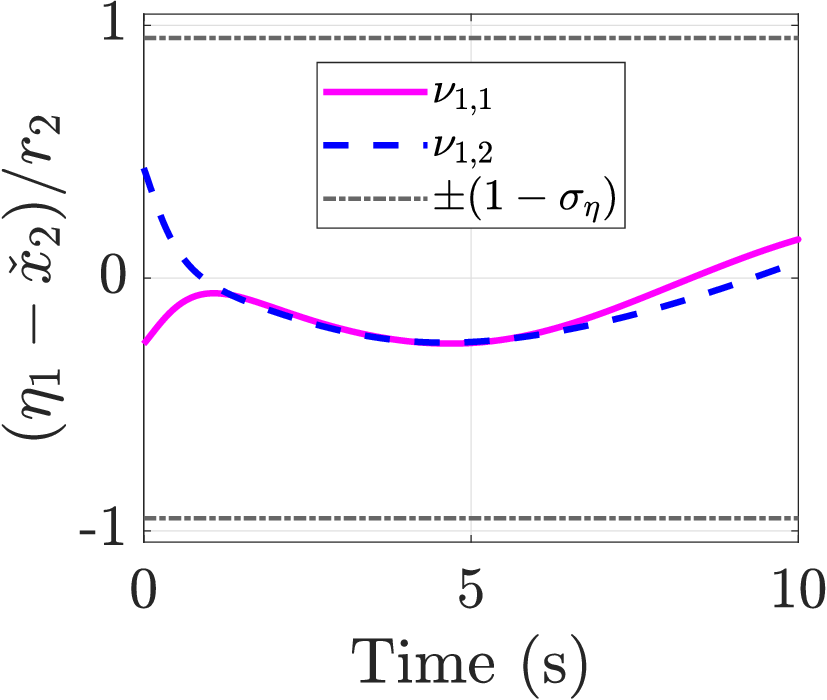}
\caption{Normalized virtual controls.}
\label{fig:lyap_nu}
\end{subfigure}%
\caption{Full-state constrained tracking under an oscillating actuator set.}
\label{fig:states}
\end{figure*}

The performance of the proposed strategy is presented in \Cref{fig:states}. It can be observed from \Cref{fig:states_x1,fig:lyap_err} that both output states converge to their respective reference trajectory within a brief transient of approximately $1.5$ s. During this transient, the controller generates a larger commanded input to nullify the tracking errors. As shown in \Cref{fig:states_uc}, these commands decrease to near zero once the tracking errors have converged to zero. Correspondingly, the realized inputs in \Cref{fig:states_u} settle near zero and remain strictly within the actuator set throughout the time, despite the time-varying actuator bounds and the changes in the width. Furthermore, \Cref{fig:states_x1,fig:states_x2} show that both state blocks remain strictly inside their respective moving envelopes, with relative margins of $0.302$ and $0.252$, respectively. The relative margin of the realized input is $0.327$. One may observe from \Cref{fig:lyap_nu} that the normalized virtual control remains well within the separation bound $\pm(1-\sigma_{\eta})=\pm0.95$. This, in turn, verifies that the APC compatibility condition is satisfied throughout the simulation.

We now investigate the effect of the control gains on the performance of the proposed APC framework. The gains are chosen from the set $k_{1}=k_{2}=k\in\{1.5,2,2.5,3\}$ with $k_{3}=4$, while the other simulation settings are kept the same to those used in \Cref{fig:states}. The performance of the proposed strategy is summarized in \Cref{tab:sweep}. It can be observed that the state and input constraints are satisfied in every case. Furthermore, as the gain increases from $1.5$ to $3$, the peak of the normalized virtual control increases almost linearly from $0.315$ to $0.675$, while remaining below the separation bound $1-\sigma_{\eta}=0.95$. Thus, the separation condition is satisfied throughout the gain variations. On the other hand, the input margin decreases from $0.383$ to $0.014$, while the peak commanded input increases from $0.8$ to $31.3$. This is primarily due to the fact that the numerator of the commanded input contains the term $\dot{\eta}_{n}$, which increases with the controller gains, as discussed in \Cref{rem:compat}. 
\begin{table}[!ht]
\centering
\caption{Effect of the control gains.}
\label{tab:sweep}
\footnotesize
\setlength{\tabcolsep}{4pt}
\begin{tabular}{cccccc}
\toprule
$k$ & $\max\nrm{(\eta_{1}-\check{x}_{2})/r_{2}}_{\infty}$ & $\mathcal{M}_{x_{1}}$ & $\mathcal{M}_{x_{2}}$ & $\mathcal{M}_{u}$ & $\max\nrm{u_{c}}_{\infty}$\\
\midrule
1.5 & 0.315 & 0.302 & 0.312 & 0.383 & 0.8\\
2.0 & 0.435 & 0.302 & 0.252 & 0.327 & 1.7\\
2.5 & 0.555 & 0.302 & 0.192 & 0.189 & 3.5\\
3.0 & 0.675 & 0.302 & 0.132 & 0.014 & 31.3\\
\bottomrule
\end{tabular}
\end{table}

We next consider a scenario where the actuator set is contracting, which may arise from actuator derating or a fault that reduces the available actuator authority. The actuator bounds are chosen as $u_{\max}(t)=[1+2e^{-at}]\mathbf{1}$ and $u_{\min}(t)=-[0.8+1.7e^{-at}]\mathbf{1}$, such that the admissible interval contracts from $(-2.5,3)$ to $(-0.8,1)$ at a rate governed by $a$. The magnitude of the lower bound contracts faster in logarithmic terms, with $\sup_{t}\lvert\mathrm{d}\ln\abs{u_{\min,l}}/\mathrm{d}t\rvert=0.68a$, attained at $t=0$.  We consider $a\in\{1,2,3,3.5\}$, for which the maximum contraction rate remains below $p_{1}p_{2}=2.5$. Thus, \Cref{ass:actuator} is satisfied, with $\varepsilon_{0}=0.12$ for $a=3.5$. The initial input is chosen as $u(0)=[0.406,-0.710]^{\top}$, while all other simulation settings are kept the same as in \Cref{fig:states}.
\begin{table}[!ht]
\centering
\caption{Effect of the contraction rate.}
\label{tab:contract}
\footnotesize
\begin{tabular}{cccc}
\toprule
$a$ & $\sup_{t}\lvert\mathrm{d}\ln\abs{u_{\min}}/\mathrm{d}t\rvert$ & $\mathcal{M}_{u}$&$\max\nrm{u_{c}}_{\infty}$\\
\midrule
1.0 & 0.680 & 0.194&1.72\\
2.0 & 1.360 & 0.142&1.72\\
3.0 & 2.040 & 0.084&1.72\\
3.5 & 2.380 & 0.056&1.72\\
\bottomrule
\end{tabular}
\end{table}
\begin{figure*}[!ht]
\centering
\begin{subfigure}{0.32\linewidth}
\centering
\includegraphics[width=\linewidth]{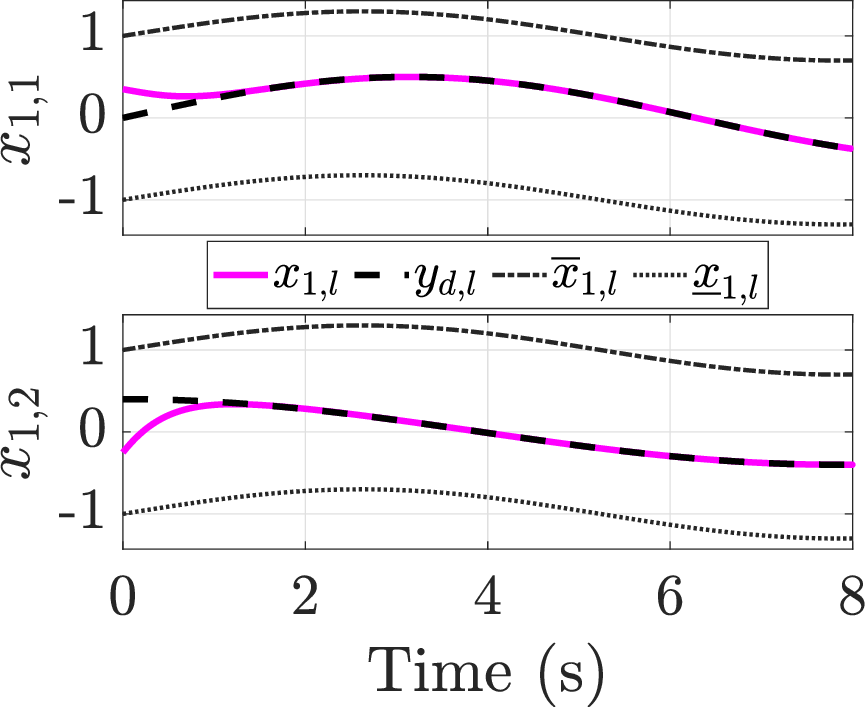}
\caption{First state block.}
\label{fig:contract_x1}
\end{subfigure}%
\hfill
\begin{subfigure}{0.32\linewidth}
\centering
\includegraphics[width=\linewidth]{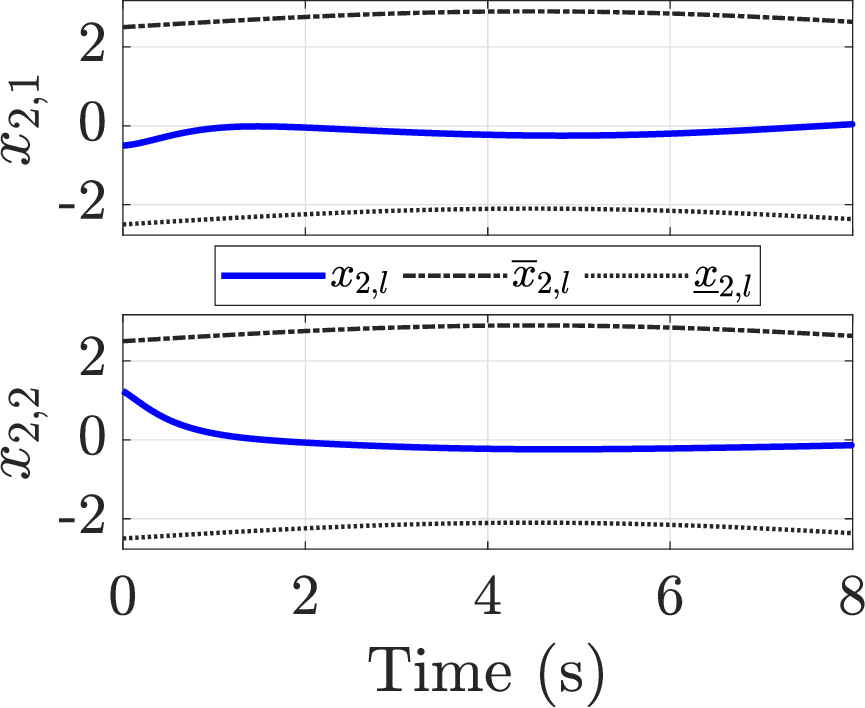}
\caption{Second state block.}
\label{fig:contract_x2}
\end{subfigure}%
\hfill
\begin{subfigure}{0.32\linewidth}
\centering
\includegraphics[width=\linewidth]{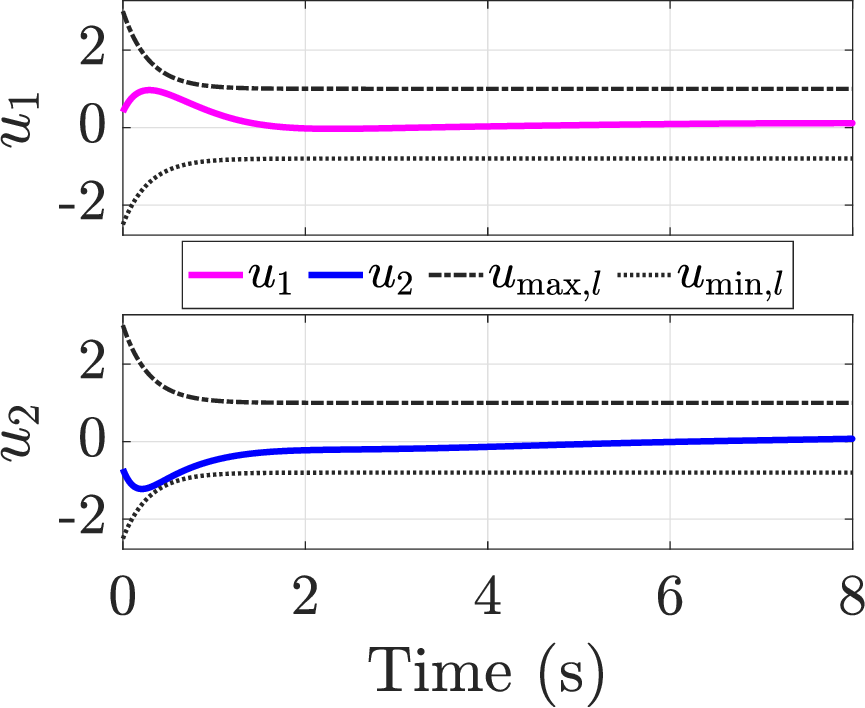}
\caption{Control inputs.}
\label{fig:contract_u}
\end{subfigure}\\
\begin{subfigure}{0.32\linewidth}
\centering
\includegraphics[width=\linewidth]{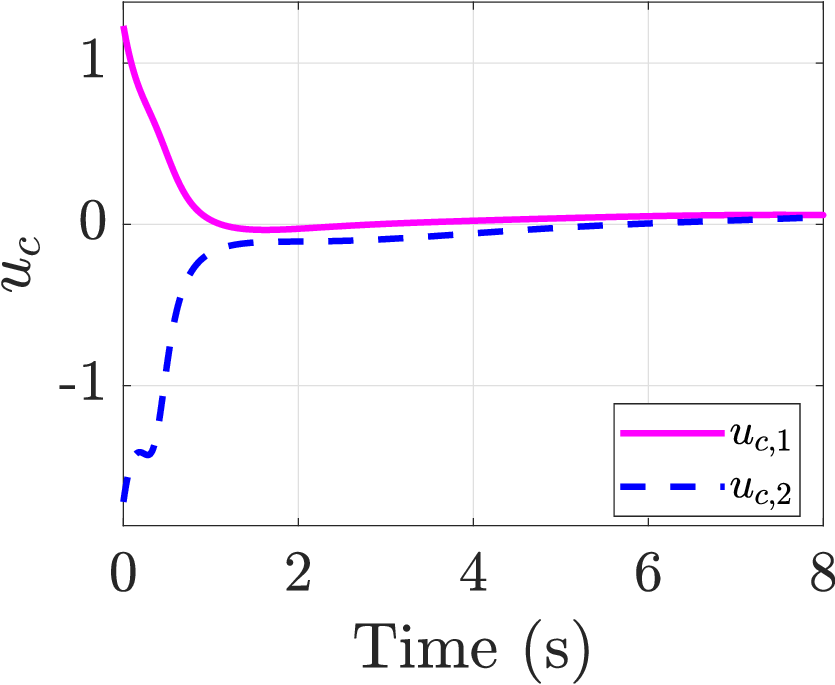}
\caption{Commanded inputs.}
\label{fig:contract_uc}
\end{subfigure}%
\hfill
\begin{subfigure}{0.32\linewidth}
\centering
\includegraphics[width=\linewidth]{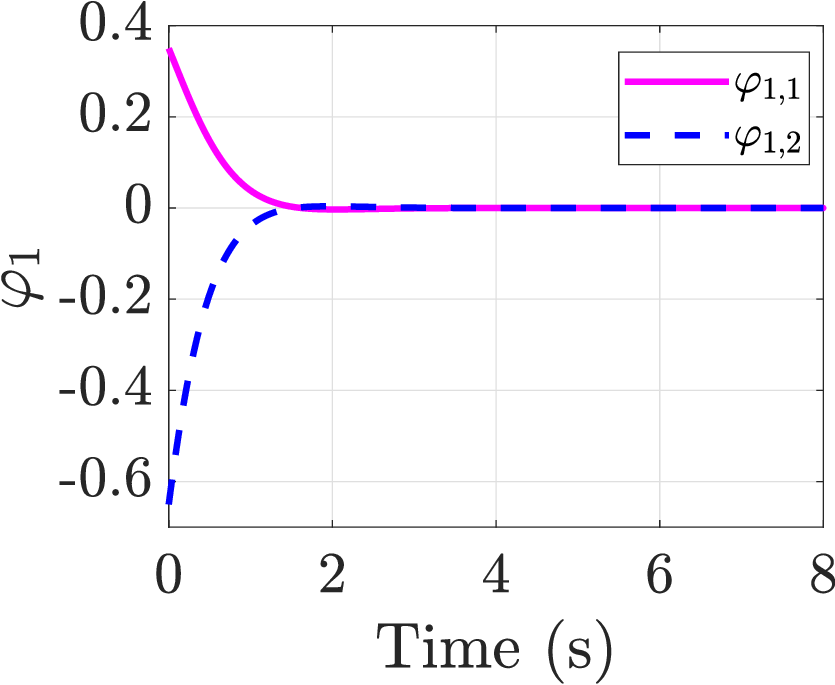}
\caption{Tracking errors.}
\label{fig:contract_err}
\end{subfigure}%
\hfill
\begin{subfigure}{0.32\linewidth}
\centering
\includegraphics[width=\linewidth]{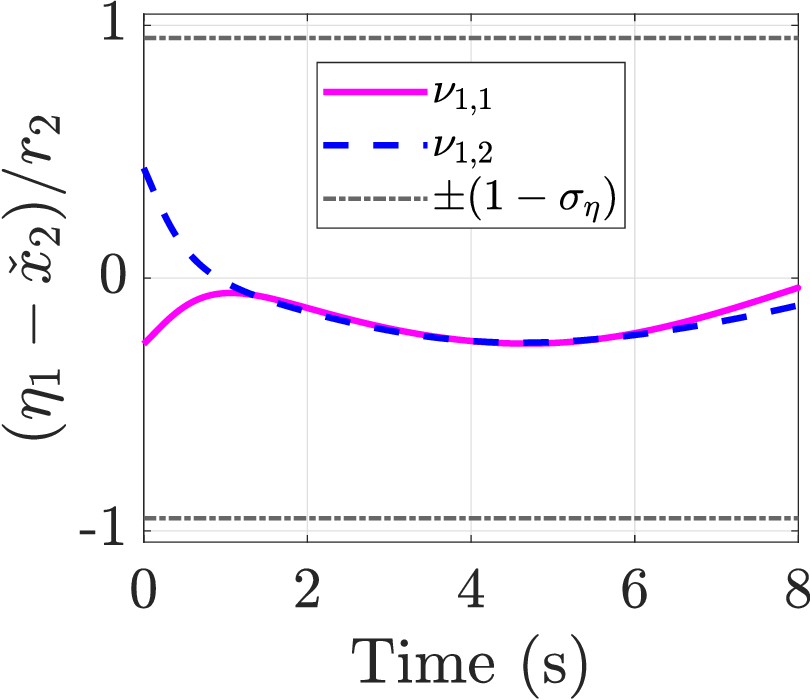}
\caption{Normalized virtual controls.}
\label{fig:contract_nu}
\end{subfigure}%
\caption{Full-state constrained tracking under a contracting actuator set.}
\label{fig:contract}
\end{figure*}

{The performance of the proposed strategy in terms of the effect of the contraction rate is summarized in \Cref{tab:contract}, while the response for fastest contraction, $a=3.5$, is presented in \Cref{fig:contract}.} It can be observed from \Cref{tab:contract} that the input margin decreases monotonically from $0.194$ to $0.056$ as the contraction rate increases from $0.68$ to $2.38$. However, it remains positive for all values of $a$, as guaranteed by \Cref{thm:apir}. Moreover, the peak commanded input remains the same at $1.72$, which is attained at $t=0$, and the tracking error converges to zero, since the interval $(-0.8,1)$ still contains the input required to track the desired reference trajectory. Furthermore, \Cref{fig:contract_x1,fig:contract_x2,fig:contract_err,fig:contract_nu} show that even for the fastest contraction, both state blocks remain within their respective envelopes, and the tracking errors converge to zero by exhibiting a similar behavior to those in \Cref{fig:states}. Thus, the proposed framework accommodates the actuator contraction without increasing the peak command or degrading the tracking accuracy. 

Finally, we assess the sensitivity of the proposed design to the choice of weights $\mu_i$. We consider the same two scenarios: the actuator set oscillates and varies in width, and the actuator set contracts. The state and input envelopes are chosen the same as in \Cref{fig:states} and in \Cref{fig:contract} (with $a=3$ and $a=3.5$). The value of Lyapunov weights are chosen from the set $\mu_{1}\in{0.0625,0.125,0.25,0.5}$, spanning a factor of eight around $\bar{r}{1}^{2}/4=0.25$, while keeping $\mu_{2}=1.5625$ fixed. By keeping the other settings the same as those in \Cref{fig:states}, we summarize the performance in \Cref{tab:weights}. In the table, nominal means the case in which the actuator set oscillates, while $a=3$ and $a=3.5$ correspond to the contracting actuator set.
It can be observed that the state and input constraints are satisfied in each case. As $\mu_{1}$ decreases from $0.25$ to $0.0625$, the peak value of $\eta_{2}$ increases from $1.20$ to $1.36$, while the input margins change by less than seven percent and the peak command by less than two percent. In contrast, increasing $\mu_{1}$ to $0.5$ leaves the response for the case when the actuator set oscillates and varies in width practically unchanged but reduces the input margins to $0.054$ and $0.024$ for $a=3$ and $a=3.5$, respectively. This is because the cross term in \eqref{eq:etai} increases with $\mu_{1}/\mu_{2}$, making its effect on the commanded input more pronounced as the actuator set contracts. Thus, the choice $\mu_{i}=\bar{r}_{i}^{2}/4$ lies near the upper end of a broad range of acceptable weights.
\begin{table}[!ht]
\centering
\caption{Effect of the Lyapunov weights.}
\label{tab:weights}
\footnotesize
\setlength{\tabcolsep}{5pt}
\begin{tabular}{cccccc}
\toprule
 & \multicolumn{3}{c}{$\mathcal{M}_{u}$} & & \\
\cmidrule(lr){2-4}
$\mu_{1}$ & nominal & $a=3$ & $a=3.5$ & $\max\nrm{\eta_{2}}_{\infty}$ & $\max\nrm{u_{c}}_{\infty}$\\
\midrule
0.0625 & 0.306 & 0.084 & 0.058 & 1.36 & 1.70\\
0.125  & 0.314 & 0.086 & 0.059 & 1.30 & 1.68\\
0.25   & 0.327 & 0.084 & 0.056 & 1.20 & 1.67\\
0.5    & 0.322 & 0.054 & 0.024 & 1.10 & 1.74\\
\bottomrule
\end{tabular}
\end{table}
\section{Conclusions}\label{sec:conclusions}
This work developed a time-varying APC framework for block strict-feedback nonlinear systems subject to time-varying asymmetric actuator and full state constraints. The proposed APIR keeps the realized input inside a moving actuator set whenever each bound contracts more slowly than the decay rate of the realization, with a closed-form lower bound on its gain. By applying the logarithmic barrier coordinate at every block of the backstepping recursion, the full state constraints hold without barrier widths reserved for the errors, and the feasibility condition of BLF backstepping is replaced by a separation condition on the virtual controls. Under APC compatibility, the tracking error converges exponentially to zero, while every state and the realized input remain inside their moving sets for all time.
\bibliographystyle{IEEEtran}
\bibliography{ref_TVBS_APC}

\end{document}